\documentclass{article}%
\usepackage{amsmath}
\usepackage{amsfonts}
\usepackage{amssymb}
\usepackage{graphicx}%
\providecommand{\U}[1]{\protect\rule{.1in}{.1in}}

\begin{document}

\date{}

\title{{\Large \textbf{Modeling and Simulation of Liquid Crystal Elastomers}}}
\author{Wei Zhu\thanks{Department of Mathematics, University of Alabama, Box 870350, Tuscaloosa,
AL 35487. Email: wzhu7@bama.ua.edu. This work was done when the first author was at the Courant Institute
of Mathematical Sciences, New York University, 251 Mercer Street, New York, NY, 10012. }
, Michael Shelley\thanks{Courant Institute of Mathematical Sciences, New York
University, 251 Mercer Street, New York, NY, 10012. E-mail:
shelley@cims.nyu.edu}, and Peter Palffy-Muhoray\thanks{Liquid Crystal
Institute, Kent Stat University, Kent, Ohio 44242. E-mail:
mpalffy@cpip.kent.edu} }
\maketitle

\section*{\centering Abstract}

\emph{We consider a continuum model describing the dynamic behavior of nematic
liquid crystal elastomers (LCEs) and implement a numerical scheme to solve the
governing equations. In the model, the Helmholtz free energy and Rayleigh
dissipation are used, within a Lagrangian framework, to obtain the equations of
motion. The free energy consists of both elastic and liquid crystalline
contributions, each of which is a function of the material displacement and
the orientational order parameter. The model gives dynamics for the material
displacement, the scalar order parameter and the nematic director, the latter
two of which correspond to the orientational order parameter tensor. Our
simulations are carried out by solving the governing equations using an
implicit-explicit scheme and the Chebyshev polynomial method. The simulations
show that the model can successfully capture the shape changing dynamics of
LCEs that have been observed in experiments, and also track the evolution of
the order parameter tensor.}

\section{Introduction}

Liquid crystal elastomers (LCEs) are orientationally ordered solids, combining
features of liquid crystals and elastic solids. \ They were first proposed by
de Gennes \cite{deGennes75} and first synthesized by Finkelmann et al
\cite{FKRM81}. \ They consist of weakly cross-linked liquid crystal polymers
with orientationally ordered side or main-chain mesogenic units. They exhibit
many new phenomena not found in either liquid crystals or polymers. \ The
salient feature of LCEs is the strong coupling between mechanical deformation
and orientational order. As a consequence of this coupling, mechanical strains
change the order parameter and hence physical properties of LCEs, and,
conversely, extermal stimuli, such as light, affecting orientational
order\ can produce large shape changes \cite{BP94, PCFS04, WT03, YNI03}.

Although many fascinating experimental results have been obtained studying the dynamic response of LCEs
to external stimuli \cite{CTT02, CTTW01, FNPW01, PCFS04, PCMTM06, THCW03, WKTNVSGTB07, YNI03},
their dynamics is not fully understood. In this paper, we implement a non-local continuum model
\cite{EMPS06}, chose and explicitly define a specific representation and carry
out numerical simulations to explore the dynamic behavior. \ Our work thus
includes both components of modeling and simulation.

In the model, the Helmholtz free energy and Rayleigh dissipation are combined,
using a Lagrangian approach, to obtain the dynamics. \ The free energy
consists of both elastic and nematic contributions, and includes volume
conservation. As a special case of the continuum model, we choose a simple
local form of the nematic free energy, the Maier-Sauper free energy, to
describe nematic contributions. Our model considers only the uniaxial phase of
nematic LCEs, for which the order parameter tensor can be expressed in terms
of a scalar order parameter and nematic director; direct contributions to the
free energy from spatial variations of the order parameter and director are
neglected. These simplifications make our model more tractable both
theoretically and numerically. \ Subsequently, the governing equations can be
derived explicitly using both conserved and non-conserved order parameter
dynamics.\ We thus obtain the time dependent equations for the displacement,
scalar order parameter and nematic director.

The equations obtained are more complicated than the standard Navier-Stokes
equations in the Eulerian frame. First, besides the pressure term and the
viscous term, there is also an elastic term in the velocity equation.
\ Second, our equations are written in a Lagrangian frame. This choice is
straightforward for capturing the orbit as well as the dynamics of each
particle in the LCE sample. Indeed, Eulerian coordinates are not well suited
to our problem since the domain occupied by the LCE sample varies in time.
\ Moreover, the derived velocity equation is very stiff due to the presence of
different time scales in the problem, posing challenges for the simulation.
The simulation is therefore a fascinating but a very formidable problem. \ In
this work, we employ the Chebyshev polynomial method \cite{Peyret02} to
discretize the spatial derivatives in the dynamical equations. This method, as
a typical spectral method, can achieve high accuracy, and is particularly well
suited to our simulation as our system is non-periodic. We also apply the
popular implicit-explicit (IMEX) schemes for the time-discretization of the
equations. \ Specifically, a combination of second-order Adams-Bashforth
method for explicit terms and Crank-Nicolson method for implicit terms
\cite{ARW95, GKO95, Peyret02} are used.

The paper is organized as follows. In Section 2, we give the details of the
model as well as the derivation of the governing equations. To obtain the
equations, we first calculate the functional derives of those functionals with
respect to the principal variables, i.e., material displacement, order
parameter and nematic director, and then apply the appropriate
conserved/non-conserved dynamics for each of the variables. In Section 3, we
present the numerics for solving the equations and then show the results of
our simulations. Conclusions are given in Section 4.

\section{Modeling nematic LCEs}

To describe the dynamics of LCEs, in addition to orientational order, one
needs to track the time evolution of the position of the crosslinks of the LCE
network. In the case of uniaxial nematic LCEs, the sample can be characterized
by the displacement, order parameter and nematic direction at each Lagrangian
lattice site, corresponding to a cross-link. \ Our work is to study how these
key variables evolve in time when the sample is subjected to external stimuli.
\ To this end, we represent the continuum model in terms of these variables,
derive the governing equations and implement the simulation by solving the
equations numerically.

Our continuum model consists of the elastic free energy density and the
nematic free energy density with coupling between orientational order and
deformation of the network, a\ Rayleigh dissipation function and a volume
preserving functional. The governing equations are derived from these by
applying the appropriate dynamics for each key variable. In what follows, the
functionals and the functional derivatives are discussed.

\subsection{Energy functionals}

The free energy in our model is composed of elastic and nematic contributions.
The elastic free energy describes the nonlocal interaction between connected
cross-links of elastomers, while the nematic free energy represents the
anisotropic dispersion interactions of the mesogenic constituents.

\subsubsection{The elastic free energy}

Let $\mbox{{\boldmath$\alpha$}}$ denote a material point in the LCEs sample, and $\mathbf{x}(\mbox{{\boldmath$\alpha$}},t)$ the
location of the point $\mbox{{\boldmath$\alpha$}}$ at time $t$. To describe the nematic ordering
in the LCEs crossing-link network, an effective dimensionless step length
tensor $\mathbf{L}$ is introduced, being written as $\mathbf{L}=\mathbf{I}%
+2\mu\mathbf{Q}$, where $\mathbf{Q}$ is the orientational order parameter
tensor, $\mathbf{I}$ is the identity matrix and $\mu$ is the relaxation
parameter. In the uniaxial phase case, $\mathbf{Q}=S(\frac{3}{2}%
\mathbf{n}\mathbf{n}^{T}-\frac{1}{2}\mathbf{I})$, where
$\mathbf{n}$ represents the unit vector along the average alignment
direction of the molecular symmetry axes, $S$ is the scalar order parameter
describing the degree of alignment of the molecular axes with $\mathbf{n}$ \cite{deGP93}.

In the undeformed state, the probability density of finding in the LCEs sample
a polymer chain of length $\mathcal{L}$ starting at $\mbox{{\boldmath$\alpha$}}$ and ending at
$\mbox{{\boldmath$\alpha$}}^{\prime}$ can be written as
\[
P_{0}(\mbox{{\boldmath$\alpha$}},\mbox{{\boldmath$\alpha$}}^{\prime})=(\frac{3}{2\pi\mathcal{L}b})^{3/2}%
(\det\mathbf{L}_{0})^{-1/2}\exp\left(-\frac{3(\mbox{{\boldmath$\alpha$}}^{\prime}-\mbox{{\boldmath$\alpha$}})^{T}%
\mathbf{L}_{0}^{-1}(\mbox{{\boldmath$\alpha$}}^{\prime}-\mbox{{\boldmath$\alpha$}})}{2\mathcal{L}b}\right),
\]
where $\mathbf{L}_{0}=\mathbf{I}+2\mu\mathbf{Q}_{0}$ is the effective step
length at the initial state \cite{WT03}. \ Since $\mathbf{Q}_{0}$ is assumed to be slowly
varying compared to the distance between crosslinks, we evaluate
$\mathbf{Q}_{0}\mathbf{\,}$\ and $\mathbf{L}_{0}$ at position $\mbox{{\boldmath$\alpha$}}$.

At time $t$, the probability density of finding a polymer ending at
$\mathbf{x}(\mbox{{\boldmath$\alpha$}},t)$ and $\mathbf{x}(\mbox{{\boldmath$\alpha$}}^{\prime},t)$ shares the same form as $P_{0}%
(\mbox{{\boldmath$\alpha$}},\mbox{{\boldmath$\alpha$}}^{\prime})$ with $\mathbf{L}_{0}(\mbox{{\boldmath$\alpha$}})$ being replaced by
$\mathbf{L}(\mbox{{\boldmath$\alpha$}},t)$. The free energy of the particular polymer initially
ending at $\mbox{{\boldmath$\alpha$}}$ and $\mbox{{\boldmath$\alpha$}}^{\prime}$ is $-kT\ln[P(\mathbf{x}(\mbox{{\boldmath$\alpha$}},t),\mathbf{x}(\mbox{{\boldmath$\alpha$}}
^{\prime},t))]$, where $k$ is the Boltzmann's constant, $T$ is the temperature.
\ The total elastic free energy at time $t$ is
\begin{align}
F_{el} &  =\frac{1}{2}\int d^{3}\mbox{{\boldmath$\alpha$}}\mathcal{F}_{el}\nonumber\\
&  =\frac{1}{2}\int d^{3}\mbox{{\boldmath$\alpha$}}\int d^{3}\mbox{{\boldmath$\alpha$}}^{\prime}\rho_{c}P_{0}%
(\mbox{{\boldmath$\alpha$}},\mbox{{\boldmath$\alpha$}}^{\prime}) \Bigl( -kT\ln[P(\mathbf{x}(\mbox{{\boldmath$\alpha$}},t),\mathbf{x}
(\mbox{{\boldmath$\alpha$}}^{\prime},t))] \Bigr) \nonumber\\
&  =\int d^{3}\mbox{{\boldmath$\alpha$}}\int d^{3}\mbox{{\boldmath$\alpha$}}^{\prime}H(\mbox{{\boldmath$\alpha$}},\mbox{{\boldmath$\alpha$}}^{\prime}%
) \Bigl( \frac{3}{2\mathcal{L}b}(\mathbf{x}(\mbox{{\boldmath$\alpha$}}^{\prime},t)-\mathbf{x}(\mbox{{\boldmath$\alpha$}},t))^{T}\mathbf{L}%
^{-1}(\mathbf{x}(\mbox{{\boldmath$\alpha$}}^{\prime},t)-\mathbf{x}(\mbox{{\boldmath$\alpha$}},t)) \nonumber\\
&+ \frac{1}{2}\ln\det\mathbf{L} \Bigr),
\end{align}
where$~\mathcal{F}_{el}$ is the elastic free energy density,
\[
H(\mbox{{\boldmath$\alpha$}},\mbox{{\boldmath$\alpha$}}^{\prime})=(\frac{1}{2}\rho_{c}kT)(\frac{3}{2\pi\mathcal{L}%
b})^{3/2}(\det\mathbf{L}_{0})^{-1/2}\exp\left(-\frac{3(\mbox{{\boldmath$\alpha$}}^{\prime}-\mbox{{\boldmath$\alpha$}}
)^{T}\mathbf{L}_{0}^{-1}(\mbox{{\boldmath$\alpha$}}^{\prime}-\mbox{{\boldmath$\alpha$}})}{2\mathcal{L}b}\right),
\]
and $\rho_{c}$ is the number density of crosslinks.

\subsubsection{Nematic free energy}

Perhaps the most successful description of nematic order is Maier-Saupe theory.
\ Here, the single particle potential is
\[
\mathcal{E}=-U\rho_{lc} S P_{2}(\cos\theta)+\frac{1}{2}U\rho_{lc}S^{2},
\]
where $U$ is an interaction strength, $\rho_{lc}$ the number density of the
liquid crystalline constituent, and $S$ the scalar parameter defined as
$S=<P_{2}(\cos\theta)>$, where $\theta$ is the angle between the symmetry axis
of a mesogen and the nematic director $\mathbf{n}$.

The nematic free energy can be written as:
\begin{align}
F_{nem} &  =\int d^{3}\mbox{{\boldmath$\alpha$}}\left(-\rho_{lc}kT \ln(\int\exp(-\frac{\mathcal{E}%
}{kT})d\Omega)\right)\nonumber\label{Nematic-energy}\\
&  =\int d^{3}\mbox{{\boldmath$\alpha$}}\left(\frac{1}{2}\rho_{lc}^{2}U S^{2}-\rho_{lc}%
kT \ln (\int\exp(\frac{S U\rho_{lc}P_{2}(\cos\theta)}{kT})d\Omega\right),
\end{align}
where $d\Omega=\sin\theta d\theta d\phi$, and $\theta$ is the polar while
$\phi$ is the azimuthal angle.

To simplify the expression for $F_{nem}$, we take the Taylor's series
expansion of the integrand, and obtain a Landau-de Gennes form for the free
energy density:
\begin{equation}
\mathcal{F}_{nem}=\frac{1}{2}C_{a}S^{2}-\frac{1}{3}C_{b}S^{3}+\frac{1}{4}%
C_{c}S^{4}+O(S^{5}),
\label{LdG}
\end{equation}
where
\[
C_{a}=\frac{1}{5}\rho_{lc}kT(\frac{\rho_{lc}U}{kT})^{2}\left(\frac{5kT}{\rho_{lc}%
U}-1\right),
\]%
\[
C_{b}=\frac{1}{35}\rho_{lc}kT(\frac{\rho_{lc}U}{kT})^{3}.
\]%
\[
C_{c}=\frac{1}{175}\rho_{lc}kT(\frac{\rho_{lc}U}{kT})^{4}.
\]
In the expressions for $C_{a}$ and $C_{b}$, $5kT/\rho_{lc}U=T/T^{\ast}\approx$
$5/6\,$, and $\rho_{lc}=\frac{200}{3}\rho_{c}$, so we write $C_{a}$ and
$C_{b}$ as
\[
C_{a}=500(\frac{T}{T^{\ast}}-1)\rho_{c}kT,
\]%
\[
C_{b}=400\rho_{c}kT.
\]%
\[
C_{c}=500\rho_{c}kT
\]
where $T^{\ast}\simeq355K$, the limit of undercooling of the isotropic phase,
is very near the nemati-isotropic transition temperature.

The nematic free energy can therefore be approximated by
\[
F_{nem}=\int d^{3}\mbox{{\boldmath$\alpha$}}\left(\frac{1}{2}C_{a}S^{2}-\frac{1}{3}C_{b}%
S^{3}+\frac{1}{4}C_{c}S^{4}\right).
\]

\subsubsection{The Rayleigh dissipation function}

The total dissipated power in the system is%
\begin{equation}
\mathcal{R}=\int d^{3}\mbox{{\boldmath$\alpha$}}\mathcal{R}_{d}
\end{equation}
and the Rayleigh dissipation function (dissipated power/volume) is
\[
\mathcal{R}_{d}=\frac{1}{2}\gamma_{1}\mathbf{D}:\mathbf{D}+\gamma_{2}%
\mathbf{D}:\dot{\mathbf{\mathbf{Q}}}+\frac{1}{2}\gamma_{3}\dot{\mathbf{Q}%
}:\dot{\mathbf{Q}},
\]
where $\mathbf{D}=(\nabla_{\mathbf{x}}\mathbf{u}+\nabla_{\mathbf{x}}\mathbf{u}^{T})/2$ is the symmetric rate-of-strain tensor, $\mathbf{u}%
=\dot{\mathbf{x}}$ and $\gamma_{i}$, $i=1,2,3$ are viscosities.

To be consistent with the variable of integration $\mbox{{\boldmath$\alpha$}}\,\ $\ in the
expressions for the free energy,  we rewrite the rate-of-strain $\mathbf{D}$
in terms of the Lagrangian coordinates. Indeed, note that the relation
\[
\nabla_{\mbox{{\boldmath$\alpha$}}}\dot{\mathbf{x}}=(\nabla_{\mathbf{x}}\mathbf{u})\mathbf{F},
\]
where $\mathbf{F}=\frac{\partial \mathbf{x}}{\partial\mbox{{\boldmath$\alpha$}}}$ is the deformation
gradient, then
\[
\mathbf{D}=\frac{1}{2}\left[ (\nabla_{\mbox{{\boldmath$\alpha$}}}\dot{\mathbf{x}})\mathbf{F}^{-1}+\mathbf{F}%
^{-T}(\nabla_{\mbox{{\boldmath$\alpha$}}}\dot{\mathbf{x}}^{T})\right].
\]

\subsubsection{Volume preserving functional}

The above free energy presents no restrictions on the sample volume. It is
known, however, from experiments, that most rubbers and LCEs are nearly volume
conserving \cite{Holzapfel00, WT03}. We therefore introduce a term controlling volume:
\[
F_{vol}=\frac{\Lambda}{2}\int d^{3}\mbox{{\boldmath$\alpha$}}(J-1)^{2},
\]
where $J=\det(\mathbf{F})$ and $\Lambda$ is a positive constant.

\subsection{Derivation of the governing equations}

The equations of motion are determined via a Lagrangian approach, by
extremizing the action in the presence of dissipation. \ The Lagrangian is
given by%
\[
\mathcal{L=}\int d^{3}\mbox{{\boldmath$\alpha$}}(\mathcal{E}_{kin}-\mathcal{F)}%
\]
where $\mathcal{E}_{kin}$ is the kinetic energy density, and $\mathcal{F\,}%
$\ is the free energy density. \ The equations of motion are given by
\begin{equation}
\int d^{3}\mbox{{\boldmath$\alpha$}}(\frac{d}{dt}\frac{\partial\mathcal{L}}{\partial\dot{\mathbf{x}}}%
-\frac{\delta\mathcal{L}}{\delta \mathbf{x}}+\frac{\delta\mathcal{R}}{\delta\dot{\mathbf{x}}%
})=0\label{eomx}%
\end{equation}%
\begin{equation}
\int d^{3}\mbox{{\boldmath$\alpha$}}(\frac{d}{dt}\frac{\partial\mathcal{L}}{\partial\dot{S}}%
-\frac{\delta\mathcal{L}}{\delta S}+\frac{\delta\mathcal{R}}{\delta\dot{S}%
})=0\label{eoms}%
\end{equation}
and%
\begin{equation}
\int d^{3}\mbox{{\boldmath$\alpha$}}(\frac{d}{dt}\frac{\partial\mathcal{L}}{\partial
\dot{\mathbf{n}}}-\frac{\delta\mathcal{L}}{\delta\mathbf{n}%
}+\frac{\delta\mathcal{R}}{\delta\dot{\mathbf{n}}})=0\label{eomn}%
\end{equation}
\bigskip

To derive the equations of motion for the displacement ($\mathbf{x}$), order parameter
($S$) and nematic director ($\mathbf{n}$), one needs to calculate the
functional directives of $F_{el}$, $F_{nem}$,
$F_{vol}$ with respect to $\mathbf{x}$, $S$, $\mathbf{n}$, and of the
kinetic energy $E_{kin}$ and dissipation\ $\mathcal{R}$ with
respect to $\dot{\mathbf{x}}$,\ $\dot{S}$ and $\dot{\mathbf{n}}$.

\subsubsection{Functional derivatives of elastic free energy}

The elastic free energy is given by
\begin{align}
F_{el}(\mbox{{\boldmath$\alpha$}}) &= \int d^{3}\mbox{{\boldmath$\alpha$}}\mathcal{F}_{el}(\mbox{{\boldmath$\alpha$}}) \nonumber \\
& = \int d^{3}\mbox{{\boldmath$\alpha$}}d^{3}\mbox{{\boldmath$\alpha$}}^{\prime}H(\mbox{{\boldmath$\alpha$}},
\mbox{{\boldmath$\alpha$}}^{\prime}) \Bigl( \frac{3}{2\mathcal{L}b}%
(\mathbf{x}(\mbox{{\boldmath$\alpha$}}^{\prime},t)-\mathbf{x}(\mbox{{\boldmath$\alpha$}},t))^{T}\mathbf{L}^{-1}
(\mathbf{x}(\mbox{{\boldmath$\alpha$}}^{\prime},t)-\mathbf{x}(\mbox{{\boldmath$\alpha$}},t))\nonumber \\
&+\frac{1}{2}\ln\det\mathbf{L} \Bigr)
\end{align}

We calculate the functional derivatives by the standard procedure, which we
illustrate here in computing $\delta F_{el}/\delta \mathbf{x}$. Let
$\mathbf{y}=\mathbf{y}(\mbox{{\boldmath$\alpha$}})$ be an arbitrary function. For convenience, we denote
$\mathbf{a}_{i}=\mathbf{x}_{i}(\mbox{{\boldmath$\alpha$}}^{\prime},t)-\mathbf{x}_{i}(\mbox{{\boldmath$\alpha$}},t)$, $\mathbf{b}_{i}=\mathbf{y}_{i}(\mbox{{\boldmath$\alpha$}}^{\prime
})-\mathbf{y}_{i}(\mbox{{\boldmath$\alpha$}})$, and $g(\mathbf{x})=(\mathbf{x}(\mbox{{\boldmath$\alpha$}}^{\prime},t)-\mathbf{x}(\mbox{{\boldmath$\alpha$}},t))^{T}%
\mathbf{L(\mbox{{\boldmath$\alpha$}},t)}^{-1}(\mathbf{x}(\mbox{{\boldmath$\alpha$}}^{\prime},t)-\mathbf{x}(\mbox{{\boldmath$\alpha$}},t))$. Then for any
$\epsilon>0$,
\begin{align}
g(\mathbf{x}+\epsilon \mathbf{y})  &  =\mathbf{L}_{ij}^{-1}(\mbox{{\boldmath$\alpha$}},t)(\mathbf{a}_{i}+\epsilon
\mathbf{b}_{i})(\mathbf{a}_{j}+\epsilon \mathbf{b}_{j})\nonumber\\
&  =g(\mathbf{x})+\epsilon\mathbf{L}_{ij}^{-1}(\mbox{{\boldmath$\alpha$}},t)(\mathbf{a}_{i}\mathbf{b}_{j}+\mathbf{a}_{j}%
\mathbf{b}_{i})+O(\epsilon^{2}).\nonumber
\end{align}

One gets
\[
lim_{\epsilon\rightarrow 0}\frac{1}{\epsilon} \Bigl( g(\mathbf{x}+\epsilon \mathbf{y})-g(\mathbf{x}) \Bigr)=\mathbf{L}%
_{ij}^{-1}(\mbox{{\boldmath$\alpha$}},t)(\mathbf{a}_{i}\mathbf{b}_{j}+\mathbf{a}_{j}\mathbf{b}_{i}),
\]
and then
\begin{align}
\frac{d}{d\epsilon}F_{el}(\mathbf{x}+\epsilon \mathbf{y})|_{\epsilon=0}
&  =\int d^{3}\mbox{{\boldmath$\alpha$}}\int d^{3}\mbox{{\boldmath$\alpha$}}^{\prime}H(\mbox{{\boldmath$\alpha$}},\mbox{{\boldmath$\alpha$}}^{\prime}%
)(\frac{3}{2\mathcal{L}b})\mathbf{L}_{ij}^{-1}(\mbox{{\boldmath$\alpha$}},t)(\mathbf{a}_{i}\mathbf{b}_{j}+\mathbf{a}_{j}%
\mathbf{b}_{i})\nonumber\\
&  =\int d^{3}\mbox{{\boldmath$\alpha$}}\int d^{3}\mbox{{\boldmath$\alpha$}}^{\prime}H(\mbox{{\boldmath$\alpha$}},\mbox{{\boldmath$\alpha$}}^{\prime}%
)(\frac{3}{2\mathcal{L}b}) \Bigl(  -2[\mathbf{L}_{ij}^{-1}(\mbox{{\boldmath$\alpha$}},t) \nonumber \\
& +\mathbf{L}_{ij}^{-1}(\mbox{{\boldmath$\alpha$}}^{\prime},t)]\mathbf{a}_{i}\mathbf{y}_{j}(\mbox{{\boldmath$\alpha$}}) \Bigr) \nonumber.
\end{align}

Therefore, the functional derivative of the elastic free energy $F_{el}$ is
\begin{align}
\frac{\delta F_{el}}{\delta \mathbf{x}}=-2\int d^{3}\mbox{{\boldmath$\alpha$}}^{\prime}%
H(\mbox{{\boldmath$\alpha$}},\mbox{{\boldmath$\alpha$}}^{\prime})(\frac{3}{2\mathcal{L}b})[\mathbf{L}^{-1}%
(\mbox{{\boldmath$\alpha$}},t)+\mathbf{L}^{-1}(\mbox{{\boldmath$\alpha$}}^{\prime},t)](\mathbf{x}(\mbox{{\boldmath$\alpha$}}^{\prime}%
,t)-\mathbf{x}(\mbox{{\boldmath$\alpha$}},t)).
\end{align}

We proceed similarly to calculate $\delta F_{el}/\delta S$, and get
\begin{align}
\frac{\delta F_{el}}{\delta S}  &  =\int d^{3}\mbox{{\boldmath$\alpha$}}^{\prime}%
H(\mbox{{\boldmath$\alpha$}},\mbox{{\boldmath$\alpha$}}^{\prime}) \Biggl[  \frac{1}{2}\frac{-6\mu S}{(1-\mu S)(1+2\mu S)} \nonumber \\
&+ \frac{3}{2\mathcal{L}b}\frac{1}{(1-\mu S)^{2}%
}(\mathbf{x}(\mbox{{\boldmath$\alpha$}})-\mathbf{x}(\mbox{{\boldmath$\alpha$}}^{\prime}))^{T} \left( \mathbf{I}-\frac{3(2\mu^{2}S^{2}+1)}{(2\mu
S+1)^{2}}\mathbf{n}\mathbf{n}^{T} \right) (\mathbf{x}(\mbox{{\boldmath$\alpha$}})-\mathbf{x}(\mbox{{\boldmath$\alpha$}}^{\prime
})) \Biggr].
\end{align}
\newline

We next find $\delta F_{el}/\delta\mathbf{n}$. Proceeding as above,
we obtain
\begin{align}
\frac{\delta F_{el}}{\delta\mathbf{n}}  &  =\int d^{3}%
\mbox{{\boldmath$\alpha$}}^{\prime}H(\mbox{{\boldmath$\alpha$}},\mbox{{\boldmath$\alpha$}}^{\prime})\frac{3}{2\mathcal{L}b}\frac{-6\mu
S}{(1-\mu S)(2\mu S+1)} \Biggl[ (\mathbf{n}\cdot(\mathbf{x}(\mbox{{\boldmath$\alpha$}})-\mathbf{x}(\mbox{{\boldmath$\alpha$}}^{\prime
})))(\mathbf{x}(\mbox{{\boldmath$\alpha$}})-\mathbf{x}(\mbox{{\boldmath$\alpha$}}^{\prime}))\nonumber\\
&  -(\mathbf{n}\cdot(\mathbf{x}(\mbox{{\boldmath$\alpha$}})-\mathbf{x}(\mbox{{\boldmath$\alpha$}}^{\prime})))^{2}\mathbf{n} \Biggr].
\end{align}

The purpose of the non-local description in the model rather than a gradient
expansion is to ensure that no artifacts arise in the dynamical equations due
to truncation of the gradient expansion when the variation is carried out \cite{OB85}.
\ Once the functional derivatives have been evaluated, \ gradient expansions
can safely be carried out. \ Since evaluation of the integrals is cumbersome
and computationally expensive, we therefore turn here long-wavelength
expansions of the integrands, and and apply these to $\delta F_{el}/\delta \mathbf{x}$,
$\delta F_{el}/\delta S$ and $\delta F_{el}/\delta\mathbf{n}$.

Let $\mbox{{\boldmath$\alpha$}}^{\prime}-\mbox{{\boldmath$\alpha$}}=(\frac{3}{2\mathcal{L}b})^{-1/2}\mathbf{L}%
_{0}^{1/2}\mbox{{\boldmath$\beta$}}$, then

\[
d^{3}\mbox{{\boldmath$\alpha$}}^{\prime}=(\frac{3}{2\mathcal{L}b}%
)^{-3/2}(\det(\mathbf{L}_{0})^{1/2}d^{3}\mbox{{\boldmath$\beta$}},
\]
\[
H(\mbox{{\boldmath$\alpha$}},\mbox{{\boldmath$\alpha$}}^{\prime})=(\frac{1}{2}\rho kT)(\frac{3}{2\pi\mathcal{L}b})^{3/2}
(\det\mathbf{L}_{0})^{-1/2}\exp(-\mbox{{\boldmath$\beta$}}\cdot\mbox{{\boldmath$\beta$}}),
\]

\[
\mathbf{L}^{-1}(\mbox{{\boldmath$\alpha$}},t)+\mathbf{L}^{-1}(\mbox{{\boldmath$\alpha$}}^{\prime},t)\simeq2\mathbf{L}
^{-1}(\mbox{{\boldmath$\alpha$}},t)+\frac{\partial\mathbf{L}^{-1}}{\partial\mbox{{\boldmath$\alpha$}}}(\frac
{3}{2\mathcal{L}b})^{-1/2}\mathbf{L}_{0}^{1/2}\mbox{{\boldmath$\beta$}},
\]
and
\begin{align}
\mathbf{x}(\mbox{{\boldmath$\alpha$}}^{\prime})-\mathbf{x}(\mbox{{\boldmath$\alpha$}}) & \simeq \frac{\partial \mathbf{x}}{\partial\mbox{{\boldmath$\alpha$}}}(\frac{3}{2\mathcal{L}b})^{-1/2} \mathbf{L}_{0}^{1/2}\mbox{{\boldmath$\beta$}} \nonumber \\
&+\frac{1}{2}\frac{\partial^{2}\mathbf{x}}
{\partial\mbox{{\boldmath$\alpha$}}_{1}\partial\mbox{{\boldmath$\alpha$}}_{2}}(\frac{3}{2\mathcal{L}b})^{-1/2}%
\mathbf{L}_{0}^{1/2}\mbox{{\boldmath$\beta$}}_{1}(\frac{3}{2\mathcal{L}b})^{-1/2}\mathbf{L}%
_{0}^{1/2}\mbox{{\boldmath$\beta$}}_{2}. \nonumber
\end{align}

To make the expression of $\delta F_{el}/\delta \mathbf{x}$ clear, we here
consider its $i^{th}$ component
\begin{align}
(\frac{\delta F_{el}}{\delta \mathbf{x}})_{i} &  =-2\pi^{-2/3}(\frac{1}{2}%
\rho_{c}kT)\int d^{3}\mbox{{\boldmath$\beta$}}\exp(-\mbox{{\boldmath$\beta$}}\cdot\mbox{{\boldmath$\beta$}})[\mathbf{L_{ij}}^{-1}%
(\mbox{{\boldmath$\alpha$}})\frac{\partial^{2}\mathbf{x}_{j}}{\partial\mbox{{\boldmath$\alpha$}}_{p}\partial\mbox{{\boldmath$\alpha$}}_{q}%
}\mathbf{L}_{0,pm}^{1/2}\mbox{{\boldmath$\beta$}}_{m}\mathbf{L}_{0,qn}^{1/2}\mbox{{\boldmath$\beta$}}_{n}\nonumber\\
&  +\frac{\partial\mathbf{L_{ij}}^{-1}(\mbox{{\boldmath$\alpha$}})}{\partial\mbox{{\boldmath$\alpha$}}_{s}}%
(\mathbf{F}\mathbf{L}_{0}^{1/2})_{sm}\mbox{{\boldmath$\beta$}}_{m}(\mathbf{F}\mathbf{L}_{0}%
^{1/2})_{jn}\mbox{{\boldmath$\beta$}}_{n}]
\end{align}
We recall the useful identities: if $M_{p}=\int_{-\infty}^{\infty}\eta
^{p}\exp(-\eta^{2})d\eta$, then $M_{0}=\sqrt{\pi}$, $\ M_{2}=M_{0}/2$, and
$M_{4}=3M_{0}/4$. \ \ Then
\begin{equation}
(\frac{\delta F_{el}}{\delta \mathbf{x}})_{i}=-(\frac{1}{2}\rho_{c}%
kT)\frac{\partial(\mathbf{L}^{-1}\mathbf{F}\mathbf{L}_{0})_{ip}}%
{\partial\mbox{{\boldmath$\alpha$}}_{p}}%
\end{equation}
or
\begin{equation}
\frac{\delta F_{el}}{\delta \mathbf{x}}=-(\frac{1}{2}\rho_{c}kT)\nabla
_{\mbox{{\boldmath$\alpha$}}}\cdot(\mathbf{L}^{-1}\mathbf{F}\mathbf{L}_{0}).
\end{equation}

Proceeding similarly,
\begin{align}
\frac{\delta F_{el}}{\delta S} &  =\pi^{-3/2}(\frac{1}{2}\rho
_{c}kT)\int d^{3}\mbox{{\boldmath$\beta$}}\exp(-\mbox{{\boldmath$\beta$}}\cdot\mbox{{\boldmath$\beta$}})\frac{1}{(1-\mu S)^{2}} \Biggl[ \mbox{{\boldmath$\beta$}}
^{T}\mathbf{L}_{0}^{1/2}\mathbf{F}^{T}\mathbf{F}\mathbf{L}_{0}^{1/2}%
\mbox{{\boldmath$\beta$}} \nonumber\\
&  -\frac{3(1+2\mu^{2}S^{2})}{(1+2\mu S)^{2}}(\mbox{{\boldmath$\beta$}}^{T}\mathbf{L}_{0}%
^{1/2}\mathbf{F}^{T}\mathbf{n})^{2} \Biggr] \nonumber \\
& +\pi^{-3/2}(\frac{1}{2}\rho
_{c}kT)\int d^{3}\mbox{{\boldmath$\beta$}}\exp(-\mbox{{\boldmath$\beta$}}\cdot\mbox{{\boldmath$\beta$}})\frac{-3\mu^{2}S}{(1-\mu S)(1+2\mu
S)}\nonumber\\
&  =(\frac{1}{2}\rho_{c}kT)\frac{1}{2(1-\mu S)^{2}} \left[ \mbox{tr}(\mathbf{F}%
\mathbf{L}_{0}\mathbf{F}^{T})-\frac{3(1+2\mu^{2}S^{2})}{(1+2\mu S)^{2}%
}\mbox{tr}(\mathbf{n}\mathbf{n}^{T}\mathbf{F}\mathbf{L}%
_{0}\mathbf{F}^{T}) \right] \nonumber\\
&  -(\frac{1}{2}\rho_{c}kT)\frac{3\mu^{2}S}{(1-\mu S)(1+2\mu S)},
\end{align}

and
\begin{align}
\frac{\delta F_{el}}{\delta\mathbf{n}} &  =(\frac{1}{2}%
\rho_{c}kT)\pi^{-3/2}\int d^{3}\mbox{{\boldmath$\beta$}}\exp(-\mbox{{\boldmath$\beta$}}\cdot\mbox{{\boldmath$\beta$}})\frac{-6\mu
S}{(1-\mu S)(2\mu S+1)} \Biggl[ (\mathbf{n}^{T}\mathbf{F}\mathbf{L}_{0}%
^{1/2}\mbox{{\boldmath$\beta$}})\mathbf{F}\mathbf{L}_{0}^{1/2}\mbox{{\boldmath$\beta$}}\nonumber\\
&  -(\mathbf{n}^{T}\mathbf{F}\mathbf{L}_{0}^{1/2}\mbox{{\boldmath$\beta$}})^{2}%
\mathbf{n} \Biggr].
\end{align}
For simplicity, we denote $\mathbf{G}=\mathbf{F}\mathbf{L}_{0}^{1/2}$. The $p$th
element of $\int d^{3}\mbox{{\boldmath$\beta$}} \exp(-|\mbox{{\boldmath$\beta$}}|^{2})(\mathbf{n}^{T}%
\mathbf{G}\mbox{{\boldmath$\beta$}})\mathbf{G}\mbox{{\boldmath$\beta$}}$ is $\frac{\pi^{3/2}}
{2}\mathbf{n}_{i}\mathbf{G}_{ij}\mathbf{G}_{pk}\delta_{jk}=\frac{\pi^{3/2}}
{2}\mathbf{G}_{pk}\mathbf{G}_{ki}^{T}\mathbf{n}_{i}$. Therefore
\[
\int d^{3}\mbox{{\boldmath$\beta$}}\exp(-\mbox{{\boldmath$\beta$}}\cdot\mbox{{\boldmath$\beta$}})(\mathbf{n}^{T}\mathbf{G}\mbox{{\boldmath$\beta$}}
)\mathbf{G}\mbox{{\boldmath$\beta$}}=\frac{\pi^{3/2}}{2}\mathbf{G}\mathbf{G}^{T}\mathbf{n}=\frac{\pi^{3/2}}%
{2}\mathbf{F}\mathbf{L}_{0}\mathbf{F}^{T}\mathbf{n}.
\]
Moreover, as $(\mathbf{n}^{T}\mathbf{F}\mathbf{L}_{0}^{1/2}\mbox{{\boldmath$\beta$}}
)^{2}=\mbox{{\boldmath$\beta$}}^{T}\mathbf{L}_{0}^{1/2}\mathbf{F}^{T}\mathbf{n}\mathbf{n}^{T}\mathbf{F}\mathbf{L}_{0}^{1/2}\mbox{{\boldmath$\beta$}}$, one gets
\[
\int d^{3}\mbox{{\boldmath$\beta$}}\exp(-\mbox{{\boldmath$\beta$}}\cdot\mbox{{\boldmath$\beta$}})(\mathbf{n}^{T}\mathbf{F}%
\mathbf{L}_{0}^{1/2}\mbox{{\boldmath$\beta$}})^{2}=\frac{\pi^{3/2}}{2}\mbox{tr}(\mathbf{n}\mathbf{n}^{T}\mathbf{F}\mathbf{L}_{0}\mathbf{F}^{T}).
\]
Consequently,
\begin{align}
\frac{\delta F_{el}}{\delta\mathbf{n}}  &  = (\frac{1}{2}%
\rho_{c} kT)\frac{-3\mu S}{(1-\mu S)(1+2\mu S)}[\mathbf{F}\mathbf{L}%
_{0}\mathbf{F}^{T}-\mbox{tr}(\mathbf{n}\mathbf{n}^{T}\mathbf{F}%
\mathbf{L}_{0}F^{T})\mathbf{I}] \mathbf{n}.
\end{align}

\subsubsection{Functional derivatives of nematic free energy}

As the nematic free energy can be approximated by
\[
F_{nem}=\int d^{3}\mbox{{\boldmath$\alpha$}} \left( \frac{1}{2}C_{a}S^{2}-\frac{1}{3}C_{b}S^{3}+\frac{1}{4}C_{c}%
S^{4} \right).
\]
The functional derivative of the free energy density with respect to $S$ is
\[
\frac{\delta F_{nem}}{\delta S}=C_{a}S-C_{b}S^{2}+C_{c}S^{3},
\]
and the functional derivatives with respect to $\mathbf{x}$ and $\mathbf{n}$
are all zero.

\subsubsection{Functional derivatives of volume preservation functional}

The volume conserving term does not depend on $\dot{\mathbf{x}}$, $S$, $\mathbf{n}$, $\dot{S}$ and $\dot{\mathbf{n}}$. \ We want to find
$\delta F_{vol}/\delta \mathbf{x}$.

Proceeding as before, we obtain
\begin{equation}
\frac{d}{d\epsilon}\Biggl[F_{vol}(\mathbf{x}+\epsilon \mathbf{y}) \Biggr] |_{\epsilon=0}=\Lambda\int d^{3}\mbox{{\boldmath$\alpha$}} \Bigl[ \mbox{tr}(\mathbf{F}^{-1}\frac{\partial \mathbf{y}}{\partial\mbox{{\boldmath$\alpha$}}
})(J-1)J \Bigr],
\end{equation}
and, after integrating by parts and requiring $J=1$ on the boundary, we obtain%
\begin{equation}
\frac{\delta F_{vol}}{\delta \mathbf{x}}=-\nabla_{\mbox{{\boldmath$\alpha$}}}\cdot(\Lambda
(J-1)J\mathbf{F}^{-T}).
\end{equation}

\subsubsection{Functional derivatives of Rayleigh dissipation}

We write the dissipation as
\begin{equation}
\mathcal{R}=\int d^{3}\mbox{{\boldmath$\alpha$}} \left( \frac{1}{2}\gamma_{1}\mathbf{D}:\mathbf{D}+\gamma
_{2}\mathbf{D}:\dot{\mathbf{\mathbf{Q}}}+\frac{1}{2}\gamma_{3}\dot{\mathbf{Q}%
}:\dot{\mathbf{Q}} \right) = \mathcal{R}_{1}+\mathcal{R}_{2}+\mathcal{R}_{3}
\end{equation}

Since $\mathcal{R}_{1}$ is independent of $\mathbf{x}$, $S$, $\mathbf{n}$,
$\dot{S}$ and $\dot{\mathbf{n}}$, we calculate the functional
derivative $\delta\mathcal{R}_{1}/\delta\dot{\mathbf{\mathbf{x}}}$. \ \ Proceeding as
before, we obtain
\begin{equation}
\frac{d}{d\epsilon} \Biggl[ \mathcal{R}_{1}(\dot{\mathbf{x}}+\epsilon \mathbf{y}) \Biggr] |_{\epsilon=0}=\gamma
_{1}\int d^{3}\mbox{{\boldmath$\alpha$}} \Bigl[ \mbox{tr}(\mathbf{D}\mathbf{F}^{-T}\nabla_{\mbox{{\boldmath$\alpha$}}}\mathbf{y}%
^{T}) \Bigr],
\end{equation}
and integration by parts gives%
\[
\frac{\delta\mathcal{R}_{1}}{\delta\dot{\mathbf{x}}}=-\gamma_{1}\nabla
_{\mbox{{\boldmath$\alpha$}}}\cdot(\mathbf{D}\mathbf{F}^{-T}).
\]
\bigskip

A calculation similar to that above shows that
\[
\frac{\delta\mathcal{R}_{2}}{\delta\dot{\mathbf{x}}}=-\gamma_{2}\nabla
_{\mbox{{\boldmath$\alpha$}}}\cdot(\dot{\mathbf{Q}}\mathbf{F}^{-T}).
\]
\bigskip

We note that $\mathbf{\dot{Q}}=\dot{S}(\frac{3}{2}\mathbf{n}\mathbf{n}^{T}-\frac{1}{2}\mathbf{I})+\frac{3}{2}S(\dot
{\mathbf{n}}\mathbf{n}^{T}+\mathbf{n}\dot
{\mathbf{n}}^{T})$. Then simple calculations yield \
\[
\frac{\delta\mathcal{R}_{2}}{\delta\dot{S}}=\gamma_{2}\mathbf{D}: \left( \frac{3}%
{2}{\mathbf{n}}\mathbf{n}^{T}-\frac{1}{2}\mathbf{I} \right).
\]

and
\[
\frac{\delta\mathcal{R}_{2}}{\delta\dot{\mathbf{n}}}=3\gamma
_{2}S \Bigl[ \mathbf{D}\mathbf{n}-(\mathbf{n}^{T}\mathbf{D}%
\mathbf{n})\mathbf{n} \Bigl].
\]
\newline

Similarly, we obtain
\[
\frac{\delta\mathcal{R}_{3}}{\delta\dot{S}}=\frac{3}{2}\gamma_{3}\dot{S},
\]
and
\[
\frac{\delta\mathcal{R}_{3}}{\delta\dot{\mathbf{n}}}=\frac{9}{2}%
\gamma_{3}S^{2}\dot{\mathbf{n}}.
\]
\newline

By combining these, we obtain the functional derivatives of the Rayleigh
dissipation as
\begin{align}
\frac{\delta\mathcal{R}}{\delta\dot{\mathbf{x}}}=-\gamma_{1}\nabla_{\mbox{{\boldmath$\alpha$}}
}\cdot(\mathbf{D}\mathbf{F}^{-T})-\gamma_{2}\nabla_{\mbox{{\boldmath$\alpha$}}}\cdot
(\dot{\mathbf{Q}}\mathbf{F}^{-T}),
\end{align}
\begin{align}
\frac{\delta\mathcal{R}}{\delta\dot{S}}=\gamma_{2}\mathbf{D}: \left( \frac{3}%
{2}{\mathbf{n}}\mathbf{n}^{T}-\frac{1}{2}\mathbf{I} \right) +\frac
{3}{2}\gamma_{3}\dot{S},
\end{align}
and%
\begin{align}
\frac{\delta\mathcal{R}}{\delta\dot{\mathbf{n}}}=3\gamma
_{2}S \Bigl[ \mathbf{D}\mathbf{n}-(\mathbf{n}^{T}\mathbf{D}%
\mathbf{n})\mathbf{n} \Bigr] +\frac{9}{2}\gamma_{3}S^{2}%
\dot{\mathbf{n}}.
\end{align}

\subsubsection{The equations of motion}

We now derive the equations of motion governing the
time evolution of the velocity $\mathbf{u}$ of the elastomer, and of
the nematic order parameter $\mathbf{Q}$, expressed in terms of $S$ and
$\mathbf{n}$.

Recall that
\[
\mathcal{L=}\int d^{3}\mbox{{\boldmath$\alpha$}}(\mathcal{E}_{kin}-\mathcal{F)}%
\]
where $\mathcal{E}_{kin}$ is the kinetic energy density, $\mathcal{F}$
is the free energy density, and
\begin{equation}
\int d^{3}\mbox{{\boldmath$\alpha$}}\left(\frac{d}{dt}\frac{\partial\mathcal{L}}{\partial\dot{\mathbf{x}}}%
-\frac{\delta\mathcal{L}}{\delta \mathbf{x}}+\frac{\delta\mathcal{R}}{\delta\dot{\mathbf{x}}}\right)=0
\end{equation}%
\begin{equation}
\int d^{3}\mbox{{\boldmath$\alpha$}}\left(\frac{d}{dt}\frac{\partial\mathcal{L}}{\partial\dot{s}}%
-\frac{\delta\mathcal{L}}{\delta s}+\frac{\delta\mathcal{R}}{\delta\dot{s}}\right)=0
\end{equation}
and%
\begin{equation}
\int d^{3}\mbox{{\boldmath$\alpha$}}\left(\frac{d}{dt}\frac{\partial\mathcal{L}}{\partial
\dot{\mathbf{n}}}-\frac{\delta\mathcal{L}}{\delta\mathbf{n}%
}+\frac{\delta\mathcal{R}}{\delta\dot{\mathbf{n}}}\right)=0.
\end{equation}
\bigskip

First the Lagrangian map $\mathbf{x}(\mathbf{\mbox{{\boldmath$\alpha$}}},t)$. Recasting the
kinetic energy from the current to the initial configuration gives:
\begin{align}
E_{kin} &  =\int_{\Omega(t)}d^{3}\mathbf{x} \mathcal{E}_{kin} \nonumber \\
& =\int_{\Omega(t)}d^{3}\mathbf{x} \left( \frac{1}{2}\rho_{m}%
(\mathbf{x},t)\mathbf{u}(\mathbf{x},t)\cdot\mathbf{u}(\mathbf{x},t) \right) \nonumber\\
&  =\int_{\Omega(0)}d^{3}\mbox{{\boldmath$\alpha$}} \left( \frac{1}{2}\rho_{m}(\mathbf{x}(\mbox{{\boldmath$\alpha$}}
,t),t)\dot{\mathbf{x}}(\mbox{{\boldmath$\alpha$}},t)\cdot\dot{\mathbf{x}}(\mbox{{\boldmath$\alpha$}},t)J(\mbox{{\boldmath$\alpha$}},t) \right) \nonumber\\
&  =\int_{\Omega(0)}d^{3}\mbox{{\boldmath$\alpha$}} \left( \frac{1}{2}\rho_{m}(\mbox{{\boldmath$\alpha$}},0)\dot
{\mathbf{x}}(\mbox{{\boldmath$\alpha$}},t)\cdot\dot{\mathbf{x}}(\mbox{{\boldmath$\alpha$}},t) \right),
\end{align}
where we use mass conservation $\rho_{m}(\mathbf{x}(\mbox{{\boldmath$\alpha$}},t),t)J(\mbox{{\boldmath$\alpha$}}
,t)=\rho_{m}(\mbox{{\boldmath$\alpha$}},0)$, with $\rho_{m}$ the mass density of the
liquid crystal elastomer. We then have
\begin{equation}
\frac{\delta E_{kin}}{\delta\dot{\mathbf{x}}}=\rho_{m}\dot{\mathbf{x}}=\rho
_{m}\mathbf{u},
\end{equation}
and subsequently
\[
\frac{\partial}{\partial t}(\rho_{m}\mathbf{u})+\frac{\delta
(F_{el}+F_{nem}+F_{vol})}{\delta \mathbf{x}}+\frac
{\delta\mathcal{R}}{\delta\dot{\mathbf{x}}}=0,
\]
or
\begin{align}
\rho_{m}\frac{\partial\mathbf{u}}{\partial t} & =(\frac{1}{2}\rho_{c}
kT)\nabla_{\mbox{{\boldmath$\alpha$}}}\cdot(\mathbf{L}^{-1}\mathbf{F}\mathbf{L}_{0})+\nabla_{\mbox{{\boldmath$\alpha$}}}
\cdot(\Lambda(J-1)J\mathbf{F}^{-T}) \nonumber \\
& +\gamma_{1}\nabla_{\mbox{{\boldmath$\alpha$}}}\cdot(\mathbf{D}\mathbf{F}^{-T})
+\gamma_{2}\nabla_{\mbox{{\boldmath$\alpha$}}}
\cdot(\dot{\mathbf{Q}}\mathbf{F}^{-T}).\label{mass}
\end{align}
We remark that the coupling of strain and orientational order, the
salient aspect of liquid crystal elastomers, is implicit in the first
term of the RHS. Since $\mathbf{L}=\mathbf{I}+2\mu\mathbf{Q}$, spatial
variations of the order parameter give rise to stresses, and in turn,
to elastomer motion.

Next, consider the dynamics of the order parameter expressed through the
variables $S$ and $\mathbf{n}$. Since the Lagrangian $\mathcal{L}$
does not depend on $\dot{S}$ or $\dot{\mathbf{n}}$, the equations
of motion give
\[
\frac{\partial(F_{el}+F_{nem}+F_{vol})}{\partial
S}+\frac{\partial\mathcal{R}}{\partial\dot{S}}=0,
\]%
\[
\frac{\partial(F_{el}+F_{nem}+F_{vol})}%
{\partial\mathbf{n}}+\frac{\partial\mathcal{R}}{\partial
\dot{\mathbf{n}}}=0,
\]
or
\begin{align}
\frac{3\gamma_{3}}{2}\frac{\partial S}{\partial t} &  =-(\frac{1}{2}\rho
_{c}kT)\frac{1}{2(1-\mu S)^{2}} \Biggl[ \mbox{tr}(\mathbf{F}\mathbf{L}_{0}\mathbf{F}%
^{T})-\frac{3(2\mu^{2}S^{2}+1)}{(2\mu S+1)^{2}}\mbox{tr}(\mathbf{n}\mathbf{n}^{T}\mathbf{F}\mathbf{L}_{0}\mathbf{F}^{T}) \Biggr]\label{order}%
\nonumber \\
& +(\frac{1}{2}\rho_{c}kT)\frac{3\mu^{2}S}{(1-\mu S)(1+2\mu S)}-\left[ C_{a}%
S-C_{b}S^{2}+C_{c}S^{3} \right] \nonumber \\
& -\gamma_{2}\mathbf{D}: \left( \frac{3}{2}{\mathbf{n}%
}\mathbf{n}^{T}-\frac{1}{2}\mathbf{I} \right),
\end{align}
and
\begin{equation}
\frac{9\gamma_{3}S^{2}}{2}\frac{\partial\mathbf{n}}{\partial t}%
=(\frac{1}{2}\rho_{c}kT)\frac{3\mu S}{(1-\mu S)(1+2\mu S)} \Biggl( \mathbf{F}%
\mathbf{L}_{0}\mathbf{F}^{T}-\mbox{tr}(\mathbf{n}\mathbf{n}%
^{T}\mathbf{F}\mathbf{L}_{0}\mathbf{F}^{T})\mathbf{I} \Biggr) \mathbf{n}.\label{director}%
\end{equation}
\newline
Eqs. (\ref{mass}), (\ref{order}), and (\ref{director}) are the
equations of motion for the nematic liquid crystal elastomer
system.

We make these equations of motion nondimensional by introducing the
following dimensionless quantities:
\[
\mathbf{u}^{\prime}=\frac{\mathbf{u}}{u},\hspace{10mm}%
\mbox{{\boldmath$\alpha$}}^{\prime}=\frac{\mbox{{\boldmath$\alpha$}}}{b},\hspace{10mm}t^{\prime}=\frac{t}{\tau},
\]
where $b$ is the step length of liquid crystal, and constants $u$ and $\tau$
are to be determined.

The equation (\ref{mass}) becomes
\begin{align}
\rho_{m}\frac{u}{\tau}\frac{\partial\mathbf{u}^{\prime}}{\partial
t^{\prime}} &  =(\frac{1}{2}\rho_{c}kT)\frac{1}{b}\nabla_{\mbox{{\boldmath$\alpha$}}^{\prime}%
}\cdot(\mathbf{L}^{-1}\mathbf{F}\mathbf{L}_{0})+\frac{1}{b}\nabla
_{\mbox{{\boldmath$\alpha$}}^{\prime}}\cdot(\Lambda(J-1)J\mathbf{F}^{-T})+\nonumber\\
&  +\gamma_{1}\frac{u}{b^{2}}\nabla_{\mbox{{\boldmath$\alpha$}}^{\prime}}\cdot \Bigl(
(\nabla_{\mbox{{\boldmath$\alpha$}}^{^{\prime}}}\mathbf{u}\mathbf{F}^{-1}+\mathbf{F}%
^{-T}\nabla_{\mbox{{\boldmath$\alpha$}}^{\prime}}\mathbf{u}^{T})\mathbf{F}^{-T} \Bigr),
\end{align}
where we let $\gamma_{2}=0$ for simplicity for the time being. Letting $u=b/\tau$ and $\tau=\gamma_{3}/\rho_{c}kT$, the above
equation reads:
\begin{align}
\lambda\frac{\partial\mathbf{u}^{\prime}}{\partial t^{\prime}} &
=\frac{1}{2}\nabla_{\mbox{{\boldmath$\alpha$}}^{\prime}}\cdot(\mathbf{L}^{-1}\mathbf{F}%
\mathbf{L}_{0})+\nabla_{\mbox{{\boldmath$\alpha$}}^{\prime}}\cdot(\Lambda^{\prime}(J-1)J\mathbf{F}%
^{-T})\nonumber\label{PDE-U}\\
&  +\frac{\gamma_{1}}{2\gamma_{3}}\nabla_{\mbox{{\boldmath$\alpha$}}^{\prime}}\cdot \Bigl(
(\nabla_{\mbox{{\boldmath$\alpha$}}^{^{\prime}}}\mathbf{u}\mathbf{F}^{-1}+\mathbf{F}%
^{-T}\nabla_{\mbox{{\boldmath$\alpha$}}^{\prime}}\mathbf{u}^{T})\mathbf{F}^{-T} \Bigr),
\end{align}
where $\lambda=\rho_{c}kT\rho_{m}b^{2}/\gamma_{3}^{2}$ and
$\Lambda^{\prime}=\Lambda/\rho_{c}kT$.

With this choice of parameters we have
\begin{align}
\frac{\partial S}{\partial t^{^{\prime}}} &  =-\frac{1}{6(1-\mu S)^{2}%
} \Biggl[ \mbox{tr}(\mathbf{F}\mathbf{L}_{0}\mathbf{F}^{T})-\frac{3(1+2\mu^{2}S^{2}%
)}{(1+2\mu S)^{2}}\mbox{tr}(\mathbf{n}\mathbf{n}^{T}%
\mathbf{F}\mathbf{L}_{0}\mathbf{F}^{T}) \Biggr] \nonumber \\
&  +\frac{\mu^{2}S}{(1-\mu S)(1+2\mu S)}+\frac{200}{3} \Bigl(-5(\frac{\mbox{Tem}}%
{360}-1)S+4S^{2}-5S^{3} \Bigr) ,\label{PDE-S}
\end{align}
where $\mbox{Tem}=\mbox{Tem}(\mbox{{\boldmath$\alpha$}}^{\prime},t^{\prime})$ is a
temperature function depending on location $\mbox{{\boldmath$\alpha$}}^{\prime}$ and time
$t^{\prime}$.
\begin{align}
\frac{\partial\mathbf{n}}{\partial t^{^{\prime}}}=\frac{\mu S}%
{3S^{2}(1-\mu S)(1+2\mu S)} \Bigl[ \mathbf{F}\mathbf{L}_{0}\mathbf{F}^{T}%
-\mbox{tr}(\mathbf{n}\mathbf{n}^{T}\mathbf{F}\mathbf{L}%
_{0}\mathbf{F}^{T})\mathbf{I} \Bigr] \mathbf{n} \label{PDE-n}.
\end{align}
This equation preserves the length of the director $\mathbf{n}$,
as required.

Finally, the deformation matrix $\mathbf{F}$ satisfies
\begin{align}
\frac{\partial\mathbf{F}}{\partial t^{^{\prime}}}=\nabla_{\mbox{{\boldmath$\alpha$}}^{\prime}
}\mathbf{u}, \label{PDE-F}
\end{align}
while (again) the Lagrangian map $\mathbf{x}$ satisfies
\begin{equation}
\frac{\partial \mathbf{x}}{\partial t^{^{\prime}}}=\mathbf{u}.
\label{Lagrangian-PDE}
\end{equation}

\section{Numerical results}

We present simulations of the dynamics of an LCE sample -- using
Equations (\ref{PDE-U}--\ref{Lagrangian-PDE}) -- when exposed to external
illumination and subject to two different boundary conditions. The LCE
sample is taken as box-shaped, as in Fig.~\ref{Saddle-shape}a. In the
first set of simulations, zero-stress boundary conditions are imposed
over the sample surface (i.e., the sample is ``free''). In the second
case, one end of the sample is anchored to a wall, with the remainder
free. In either case, gravitational loads are neglected. Numerically,
the difference between these two cases lies only in the treatment of
the velocity on one face of the sample. However, the dynamics the two
cases present is quite different, as observed in \cite{PCFS04}.

\subsection{Methods}

To discretize the equations of motion, one needs to consider suitable
schemes for approximating both spatial derivatives and time
derivatives.  here we employ the spectral Chebyshev polynomial method
to discretize spatial derivatives with high efficiency and accuracy
\cite{Peyret02}. As for the time-discretization, we use a popular
implicit-explicit scheme that is a combination of second-order
Adams-Bashforth scheme for the explicit term and Crank-Nicolson scheme
for the implicit term \cite{ARW95, GKO95, Peyret02}.

We now outline the Chebyshev polynomial method and the
implicit-explicit time-stepping method. The dynamics is simulated in
the Lagrangian domain $\Omega(0)=[-a,a]\times[-b,b]\times[-c,c]$,
which by definition is fixed in time. This is trivially mapped to the
cube $[-1,1]^3$.  This cubic domain is then discretized in each
direction on the Gauss-Lobatto points (e.g. in the first coordinate,
$\mbox{{\boldmath$\alpha$}}_{1,j}=\cos(j\pi/N),~j=0(1)N$). This allows allows spatially
dependent fields, such as $\mathbf{u}$ or $\mathbf{F}$, that are
represented discretely on these points to also be represented
efficiently, via FFT, as finite sums of Chebychev polynomials
\cite{Peyret02}. The Chebychev representation can then be used to
provide highly accurate derivative approximations upon the grid. To
illustrate in one dimension, let $u(x)$ be defined on $[-1,1]$ and
approximated by $u_{N}(x)=\sum_{p=0}^N a_p T_p(x)$ where $T_p$ is the
$p^{th}$ Chebychev polynomial. The $a_p$'s are determined by requiring
$u_N$ to interpolate $u$ at the Gauss-Lobatto points. This Chebychev
representation allows us to construct approximations to
$u^{(p)}(x)$, at the Gauss-Lobatto points, that have the form
\begin{align}
u_{N}^{(p)}(x_{i})  &  = \sum_{j=0}^{N}d^{(p)}_{ij}u_{N}(x_{j}), \hspace{5mm}i
= 0(1)N,\nonumber
\end{align}
where $[d^{(p)}_{ij}]$ is the the $p^{th}$ Chebyshev differentiation
matrix (see \cite{Peyret02}). Since our three-dimensional grid is of tensor
product form, derivatives are easily gotten by application of such matrices
along lines of constant coordinate of the discretized data.

The LCE dynamics in which we are interested takes place in the
over-damped regime, that is the ``Reynolds number'',
$\lambda\gamma_3/\gamma_1$, associated with viscous fluid damping is
very small. Hence, if we retain $\mathbf{u}_t$ in the dynamics, we
must implicitly treat the viscous damping so as to avoid extreme
constraints on the time-step that would be imposed by using an explicit
scheme. Here, we choose a popular implicit-explicit method
which is described for a typical time dependent equation:
\begin{align}
\frac{du}{dt}  &  = f(u)+\nu g(u),\nonumber
\end{align}
with $f(\cdot)$, $g(\cdot)$ being nonlinear and linear terms,
respectively. We apply a second-order Adams-Bashforth method to the nonlinear
terms, and Crank-Nicholson averaging to the linear term, or
\begin{align}
\label{IMEX}
\frac{u^{n+1}-u^{n}}{\Delta t}  &  = \frac{3}{2}f(u^{n})-\frac{1}%
{2}f(u^{n-1})+\frac{\nu}{2}[g(u^{n+1})+g(u^{n})],
\end{align}
where $\Delta t$ is the time step size and $u^{n}$ is the
approximation of $u(n~\Delta t)$.  This scheme involves solution
values on three time levels. The first time-step is taken by setting
$u^{-1}=u^0=u(0)$ (see \cite{Peyret02}).

\subsection{Treatment of interior and boundary points}

Note that the equations (\ref{PDE-S}) and (\ref{PDE-n}) involve no
spatial derivatives, and so are solved directly by the second-order
Adams-Bashforth method and without using the Chebyshev
approximation. More care must be taken with the momentum equation
(\ref{PDE-U}) as it is a source of stiffness in the numerical
treatment, and since its advancement involves boundary
conditions. This is an important issue, since distinct boundary
conditions result in completely different behaviors of the LCE
sample. In what follows, we mainly discuss how to handle the equation
for both boundary and interior points.

The three-dimensional grid using the Gauss-Labatto points is composed
of both surface and interior points. To update the velocity at the
interior points, we need to solve a large linear system gotten by
applying the implicit-explicit method for the velocity equation
(\ref{PDE-U}), coupling this to boundary conditions. For a ``free''
LCE sample, this is a condition of zero normal stress, which under
discretization of velocity gradients in the viscous stress provides a
coupling condition of the boundary velocities to the interior
velocities.

\label{IMEX}
The velocity equation (\ref{PDE-U}) can be discretized as follows:
\begin{align}
\frac{\mathbf{u}^{n+1}-\mathbf{u}^{n}}{\Delta t}  &
= \frac{3}{2} \Bigl[ \frac{1}{\lambda
}\nabla\cdot(\Lambda(J-1)J\mathbf{F}^{-T} +\frac{1}%
{2}\mathbf{L}^{-1}\mathbf{F}\mathbf{L}_{0}) \Bigr] ^{n}\nonumber\\ &
-\frac{1}{2} \Bigl[ \frac{1}{\lambda}\nabla\cdot(\Lambda(J-1)J%
\mathbf{F}^{-T} +\frac{1}{2}\mathbf{L}^{-1}\mathbf{F}\mathbf{L}_{0}%
) \Bigr] ^{n-1}\nonumber\\ &
+\frac{\gamma_{1}}{4\gamma_{3}}\nabla\cdot \Bigl[ (\nabla\mathbf{u}^{n+1}%
(\mathbf{F}^{-1})^{n} +(\mathbf{F}^{-T})^{n}(\nabla\mathbf{u}^{T}%
)^{n+1}(\mathbf{F}^{-T})^{n} \Bigr] \nonumber\\ &
+\frac{\gamma_{1}}{4\gamma_{3}}\nabla\cdot \Bigl[ (\nabla\mathbf{u}^{n}%
(\mathbf{F}^{-1})^{n-1} +(\mathbf{F}^{-T})^{n-1}(\nabla\mathbf{u}^{T}%
)^{n}(\mathbf{F}^{-T})^{n-1} \Bigr],\nonumber
\end{align}
where $[\cdot]^{n}$ represents the value at $\mbox{n}\Delta t$, for
instance, $(\mathbf{F}^{-T})^{n}$ denotes the value of
$\mathbf{F}^{-T}$ at $\mbox{n}\Delta t$.  This equation amounts to a
large linear system for the unknown velocity $\mathbf{u}^{n+1}$ at the
interior points. Couplings within the matrix arise through expansion
of spatial derivatives (that is, gradients and a tensor divergence)
through the Chebyshev expansion of the velocity. Despite the many
entries in the matrix generated through the derivatives, the matrix is
nonetheless rather sparse, and we explicitly construct the entries and
store this sparse matrix. Once the boundary conditions are
appropriately integrated, we solve this large system using the
iterative GMRES method \cite{SAAD86}.

Surface values of velocity are either additional unknowns, or are
specified as in the case of having an anchored surface where
$\mathbf{u}=0$ on that face. The former is the case of the zero stress
boundary condition. In the Lagrangian frame, using Nanson's formula
\cite{Holzapfel2000}, this boundary condition can be written as
\begin{align}
\label{bdy-condition}
\Bigl[ \Lambda({J}-1)\mathbf{I} +\frac{1}{2}
\mathbf{L}^{-1}\mathbf{F}\mathbf{L}_{0}\mathbf{F}^{T} \Bigr] \cdot {J}
\mathbf{F}^{-T}\mathbf{\nu}_{0} +\frac{\gamma_{1}}{2\gamma_{3}} \Bigl[ (\nabla
_{\mbox{{\boldmath$\alpha$}}}\mathbf{u}\mathbf{F}^{-1} \nonumber \\
+\mathbf{F}^{-T}\nabla_{\mbox{{\boldmath$\alpha$}}
}\mathbf{u}^{T}) \Bigr] \cdot{J}\mathbf{F}^{-T}\mathbf{\nu}_{0}  &  =
\mathbf{0},
\end{align}
where $\mathbf{\nu}_{0}$ denotes the outward normal unit vector to the time
invariant surface $\partial\Omega(0)=\partial\Omega$.
In our method, this vector $\nu_{0}$ takes different
values at the side, edge and corner points on the boundary of the
cubic volume of the LCE sample. Specifically, at the side points,
$\nu_{0}$ takes value from the set $\{(\pm1,0,0),
(0,\pm1,0),(0,0,\pm1)\}$; at the edge points, the set becomes
$\{\frac{1}{\sqrt2}(\pm1,\pm1,0)$, $\frac{1}{\sqrt2}(\pm1,\mp 1,0)$,
$\frac{1}{\sqrt2}(\pm1,0,\pm1)$,$\frac{1}{\sqrt2}(\pm1,0,\mp1)$,
$\frac{1}{\sqrt2}(0,\pm1,\pm1)$, $\frac{1}{\sqrt2}(0,\pm1,\mp1)\}$;
and at the corner points, the set is then
$\{\frac{1}{\sqrt3}(\pm1,\pm1,\pm1)$, $\frac
{1}{\sqrt3}(\pm1,\pm1,\mp1)$, $\frac{1}{\sqrt3}(\mp1,\pm1,\pm1)$,
$\frac {1}{\sqrt3}(\mp1,\pm1,\mp1)\}$.

We rewrite the boundary condition equation (\ref{bdy-condition}) as:
\begin{align}
\left[ (\nabla_{\mbox{{\boldmath$\alpha$}}}\mathbf{u}\mathbf{F}^{-1} +\mathbf{F}^{-T}%
\nabla_{\mbox{{\boldmath$\alpha$}}}\mathbf{u}^{T}) \right] \mathbf{F}^{-T}\mathbf{\nu}_{0}  &  =
\mathbf{A}\mathbf{\nu}_{0},\nonumber
\end{align}
where
$\mathbf{A}=-\frac{2\gamma_{3}}{\gamma_{1}}[\frac{1}{2}\mathbf{L}^{-1}
\mathbf{F}\mathbf{L}_{0}+ \Lambda(\mathbf{J}-1)J\mathbf{F} ^{-T}]$.
This boundary condition is evaluated at the $(n+1)$st time-step. Given
that all of the dynamics equations, bar that for the velocity
$\mathbf{u}$, are treated explicitly, we can consider $\mathbf{F}$ and
$\mathbf{A}$ as being considered known, and the velocity gradients as
unknowns. Gradients are either tangential to the sample surface, and
hence only couple together boundary points (upon approximation of gradients
using the Chebyshev representation), or normal to the surface and hence
couple together boundary and interior points.

This yields a closed set of equations for $\mathbf{u}^{n+1}$ at the
interior points, and on those surface upon which a zero stress boundary
condition is imposed. As said above, this system is solved via the GMRES
iterative method \cite{SAAD86}.

An important issue that should be emphasized is how to employ the GMRES method
efficiently. Indeed, note that the coefficient matrix is very sparse, we just
need to store the non-zero entries of the coefficient matrix, and also the row
and column indices of these non-zero entries. Then in the matrix-vector
multiplication, these non-zero entries will be multiplied by those elements of
the vector using the above stored column indices. This procedure can save lots
of memory and also accelerate the matrix-vector multiplication considerably.

\subsection{Simulations}

Before proceeding, we discuss the choice of dimesionless parameters.
These include the coefficient of acceleration
$\lambda=\rho_{c}kT\rho_{m}b^{2}/\gamma_{3}^{2}$, the viscosity ratio
$\gamma_{1}/\gamma_{3}$, the coefficient for volume conservation
$\Lambda$, and the anisotropy of step length $\mu$ appearing in the
tensor $\mathbf{L}$.  Taking typical values \cite{PCFS04} we have
$\lambda=O(10^{-3})$ and $\gamma_{1}/\gamma_{3}=O(10^{1-2})$. Hence,
inertial forces in the material are quite small. Ideally, we should
choose a very large value of $\Lambda$ to enforce material
incompressibility, but this imposes a severe time-step restriction in
our numerical scheme; We use $\Lambda=10^{3}$. The parameter $\mu$
lies in the range $[0,1.0]$. A large value of $\mu$ corresponds to a
large order parameter, which accelerates the deformation process of
the LCE sample.  We use $\mu=0.9$ in the simulations.

We now consider the simulated dynamics of the first case of a ``free''
LCE sample being exposed to illumination from above. In this
simulation, the sample size is $8\times8\times1$, with $N_1=32$,
$N_2=32$ and $N_3=10$ points being used in each direction,
respectively. The initial data used was
$\mathbf{u}_0\equiv\mathbf{0}$, $\mathbf{n}_0\equiv\mathbf{{\hat y}}$,
$\mathbf{X}_0\equiv\mathbf{\mbox{{\boldmath$\alpha$}}}$, and $s_0\equiv {\bar s}$, where
${\bar s}$ is the constant value found as the minimizer of the
Landau-de Gennes free energy density (\ref{LdG}) given a uniform
temperature throughout the sample corresponding to 290K (in
dimensional units). As discussed earlier, we neglect thermal diffusion
and assume that the temperature is uniform in each horizontal slice of
the sample, decreasing linearly from top (420K) to bottom (290K).

\begin{figure}[h]
\begin{center}
\includegraphics[width=5.5cm,height=5.5cm]{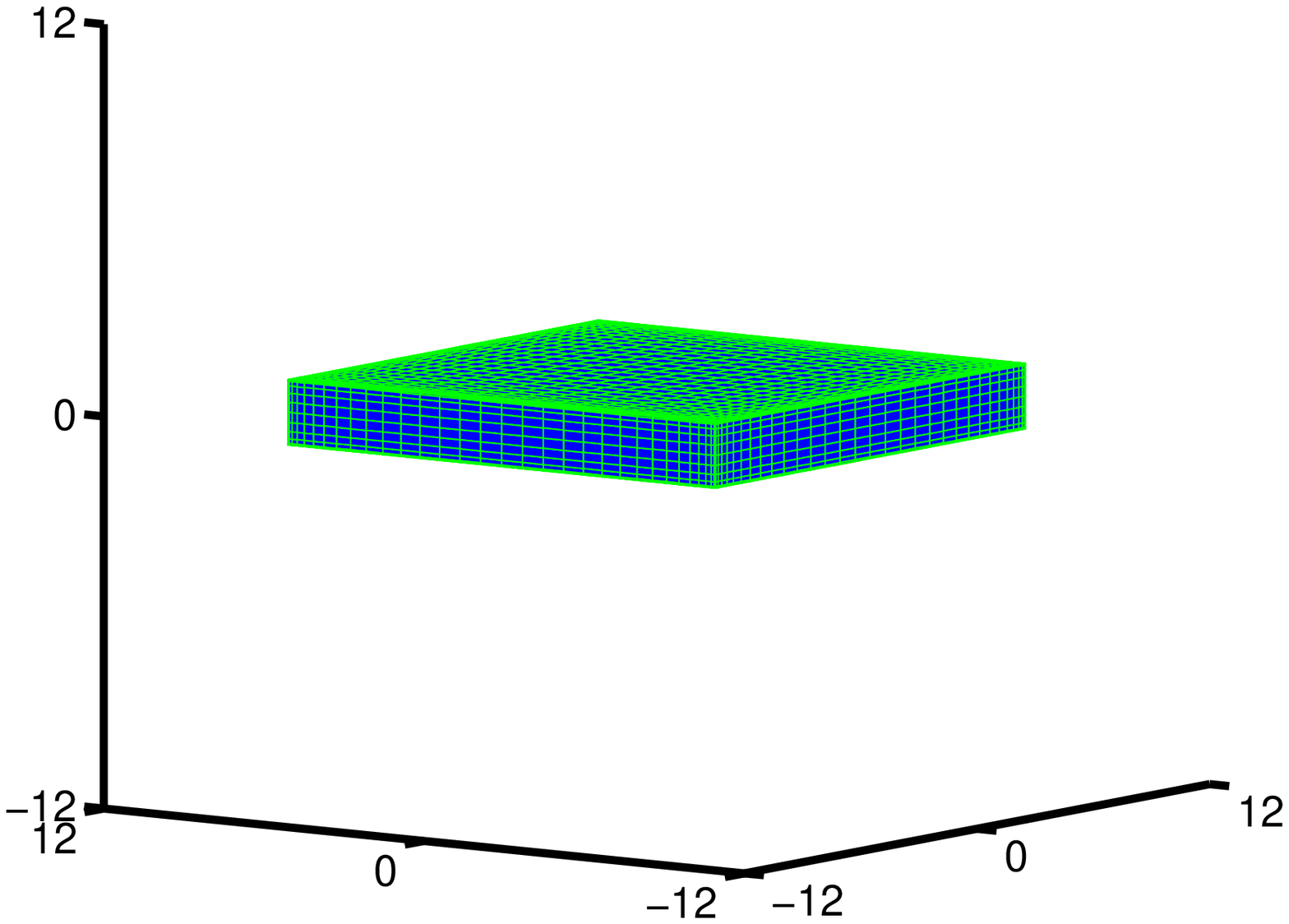}
\includegraphics[width=5.5cm,height=5.5cm]{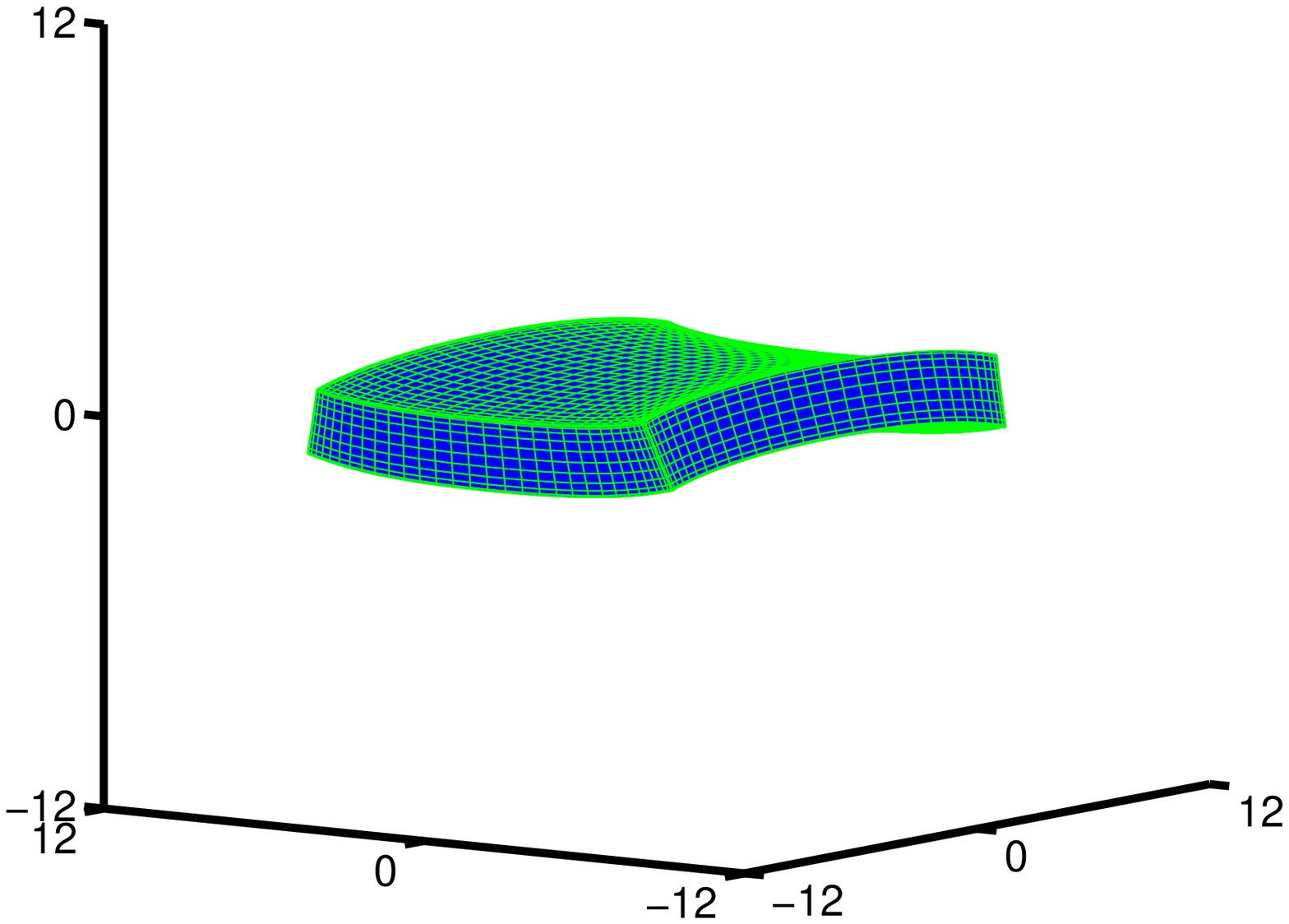}\\[0pt]($\mathbf{A}$)
\hspace{50mm} ($\mathbf{B}$)\\[0pt]%
\includegraphics[width=5.5cm,height=5.5cm]{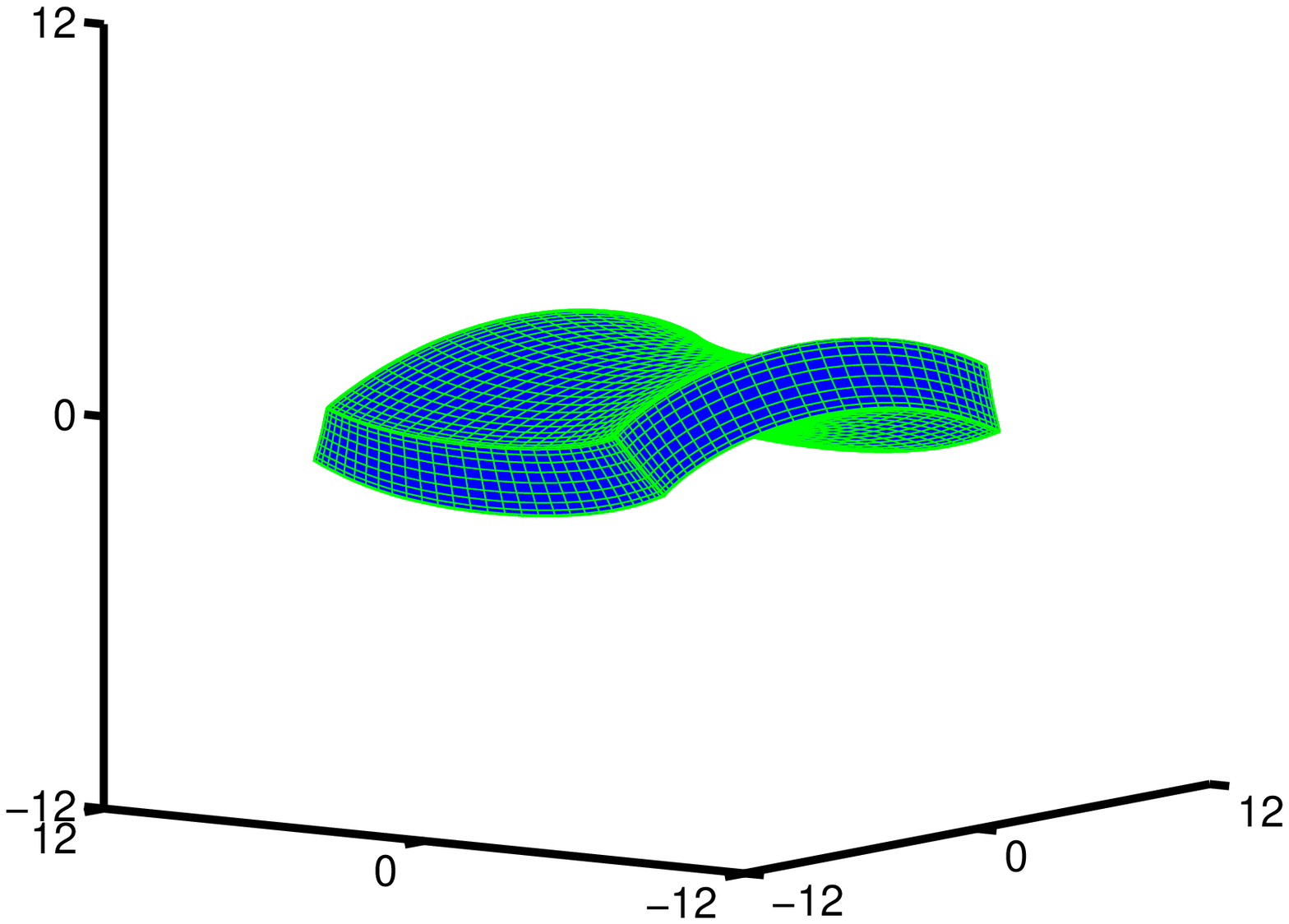}
\includegraphics[width=5.5cm,height=5.5cm]{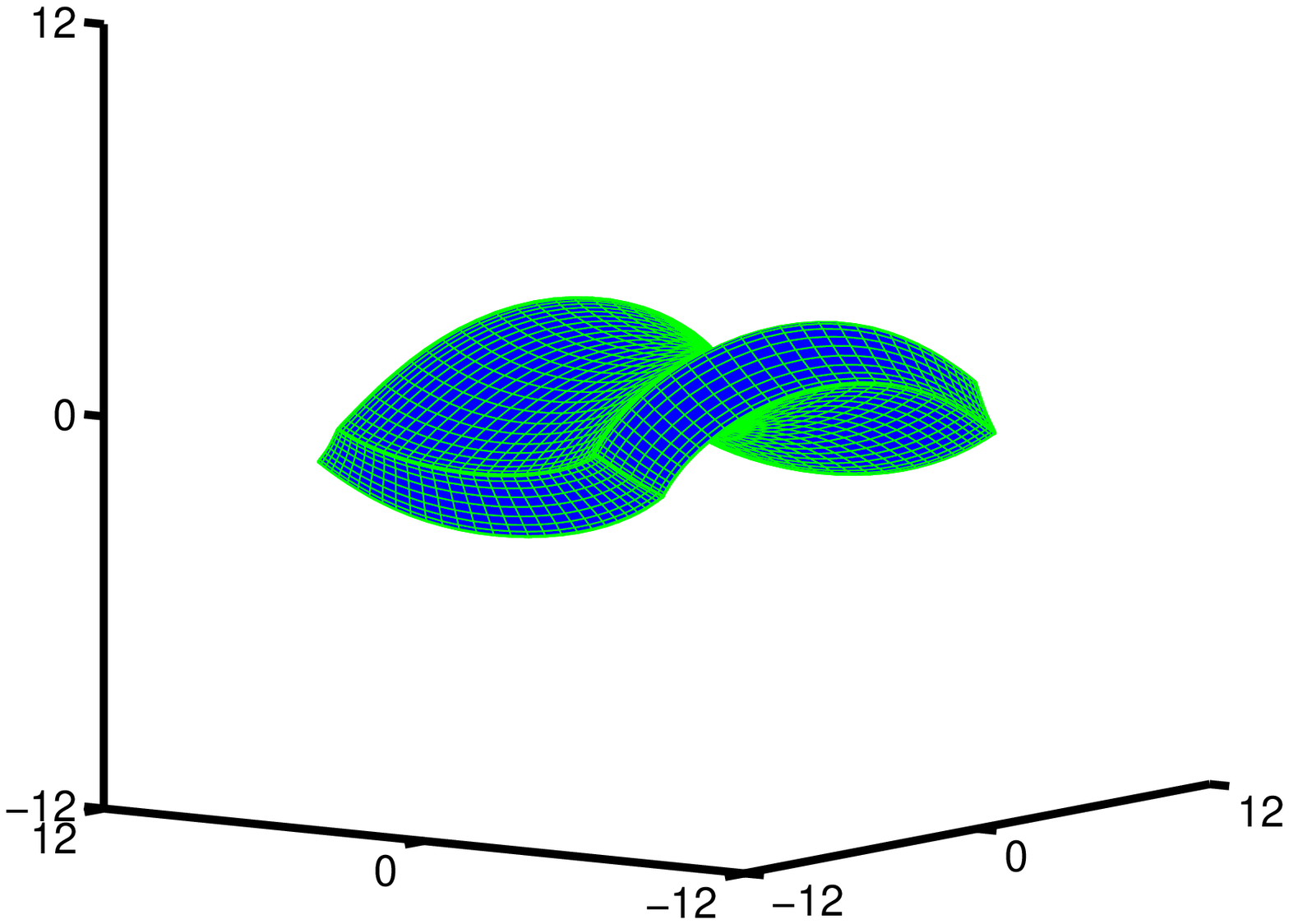}\\[0pt]($\mathbf{C}$)
\hspace{50mm} ($\mathbf{D}$)\\[0pt]%
\includegraphics[width=5.5cm,height=5.5cm]{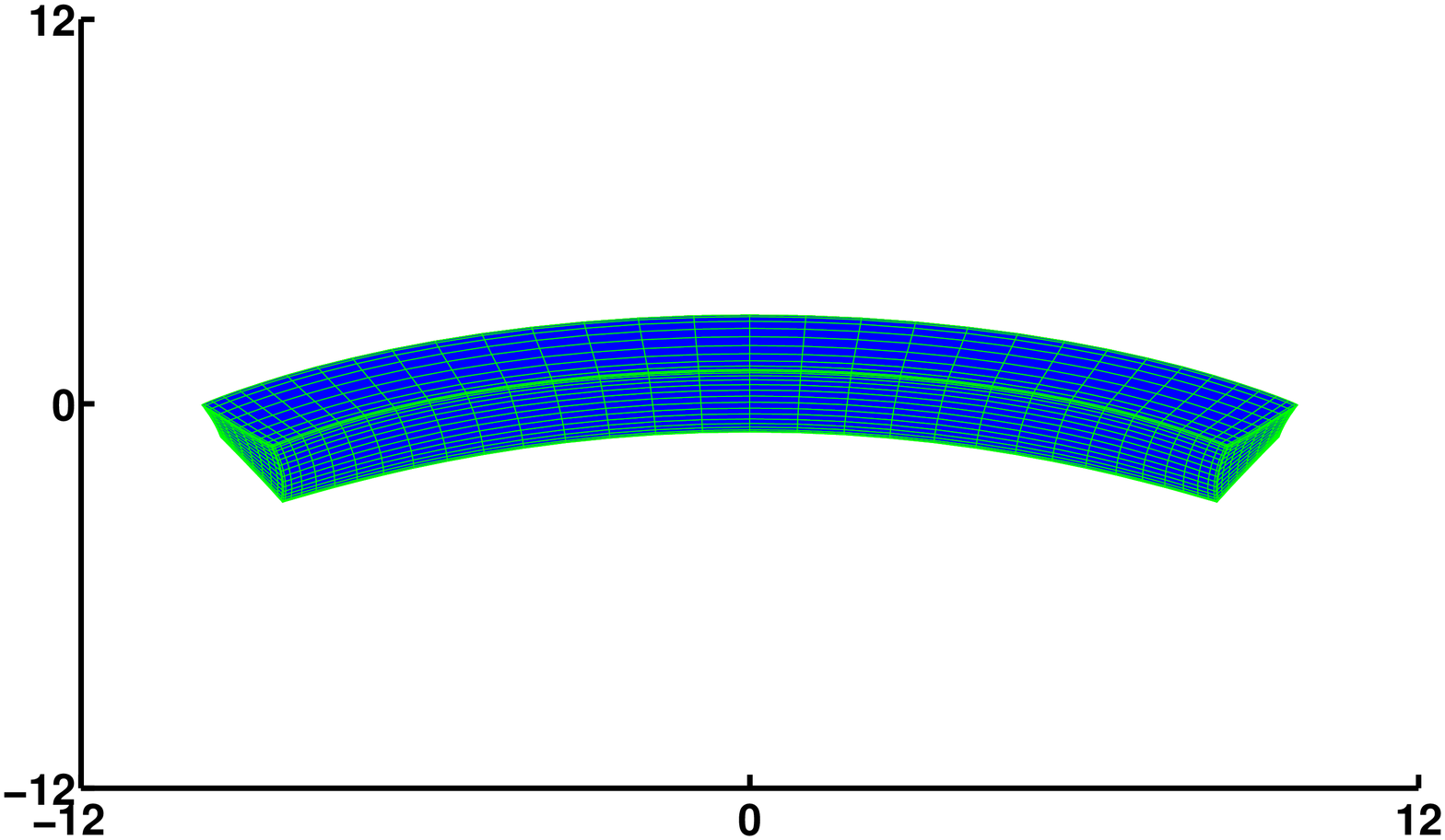}
\includegraphics[width=5.5cm,height=5.5cm]{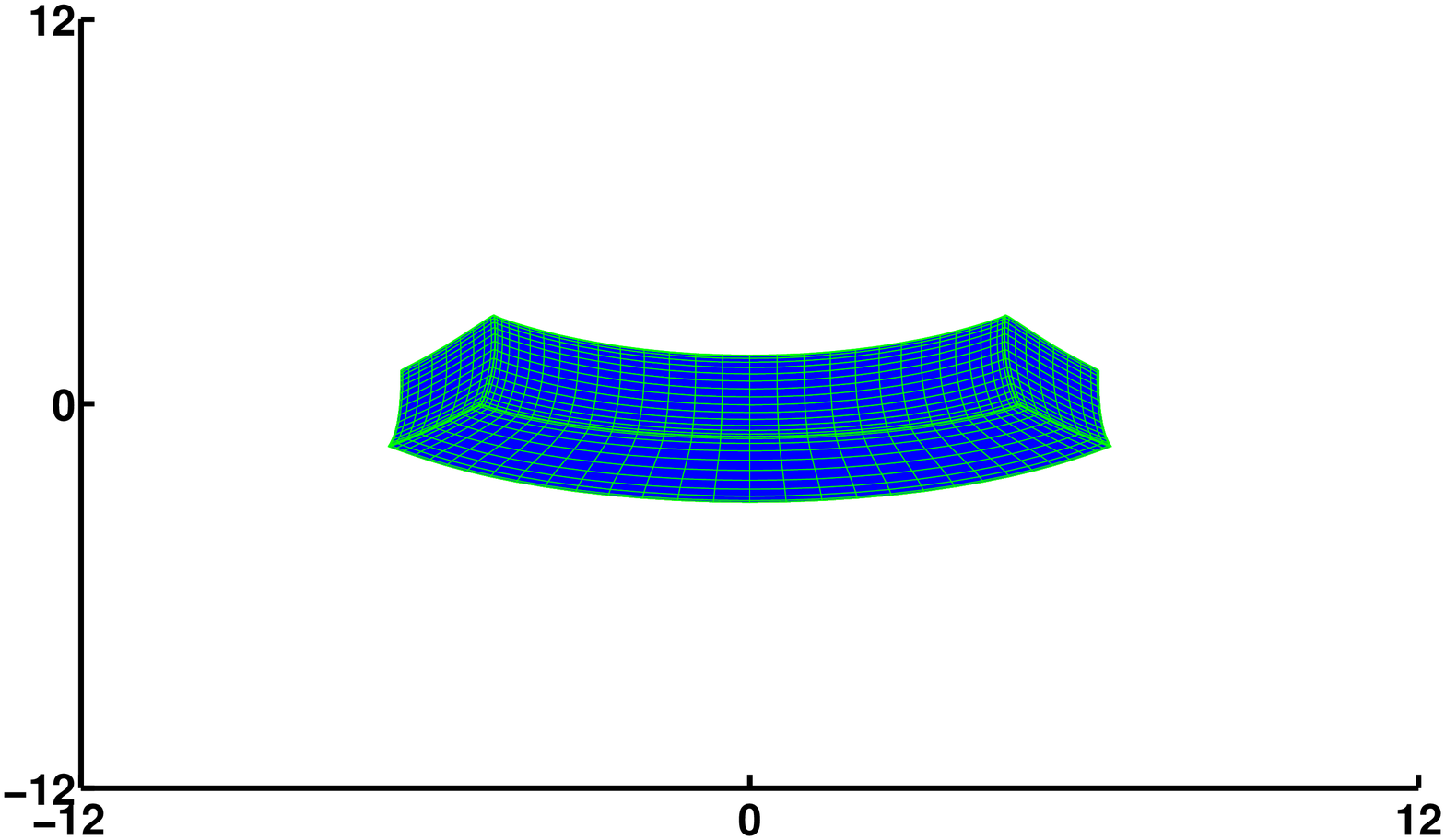}\\[0pt]($\mathbf{E}$)
\hspace{50mm} ($\mathbf{F}$)\\[0pt]
\end{center}
\caption{{\protect\small The shape evolution of the LCEs sample due to nonhomogeneous changes in temperature.
Figure $\mathbf{A}$ shows the initial state of the LCEs sample, while Figures $\mathbf{B}$, $\mathbf{C}$ are two intermediate states and
Figure $\mathbf{D}$ represents the equilibrium state. Figures $\mathbf{E}$ and $\mathbf{F}$ present the shape of the LCEs sample at the
equilibrium state (Figure $\mathbf{D}$) from two different perspectives. In this experiment, the temperature drops
linearly from the top of the LCEs sample to its bottom while distributes uniformly on each horizontal slices,
and the temperature spatial distribution is preserved during the evolution process. This numerical experiment
simulates the real one shown in Figure 4 in \cite{PCFS04}.}}%
\label{Saddle-shape}%
\end{figure}

Figure \ref{Saddle-shape} shows the deformation process from this
initial configuration. The final result -- a saddle shape -- is very
similar to that observed in the experiment of Palffy-Muhoray {\it et
al.} (see Fig. 4 of \cite{PCFS04}).  The evolution proceeds in three
stages: an initially slow and small bending, followed by rapid and
large deformation, and finally a slow relaxation to a terminal
shape. This dynamics is driven by the evolution of the orientational
order parameter, $s$, as it adjusts its values (low on the top and higher
on the bottom) in response to the imposed temperature gradient. The
inhomogeneous spatial distribution of orientational order, especially
through the thickness of the LCE sample, gives rise to large stresses
and hence creates a strong driving force towards changing the shape of
the sample.

The last two plots of Fig. \ref{Saddle-shape} show the late-time
deformed sample from two different perspectives. Here one finds that
the length along the $y-$axis has become shorter, while that along the
$x-$axis has increased. This is again due to the time evolution of
order parameter. At the top surface, given its increased temperature,
the degree of order of the rod-like mesogens decreases. Since these
mesogens are initially aligned along the $y-$direction, this loss of
order leads to a contraction of the sample along the $y-$direction and
corresponding extensions along the $x-$ and $z-$directions. Since the
temperature is different on each horizontal layer, the degree of
contraction also is also different. It is this difference in
contraction and expansion through the thickness of LCE sample that
results in the observed saddle-shaped deformation.

\begin{figure}[h]
\begin{center}
\includegraphics[width=6.0cm,height=6.0cm]{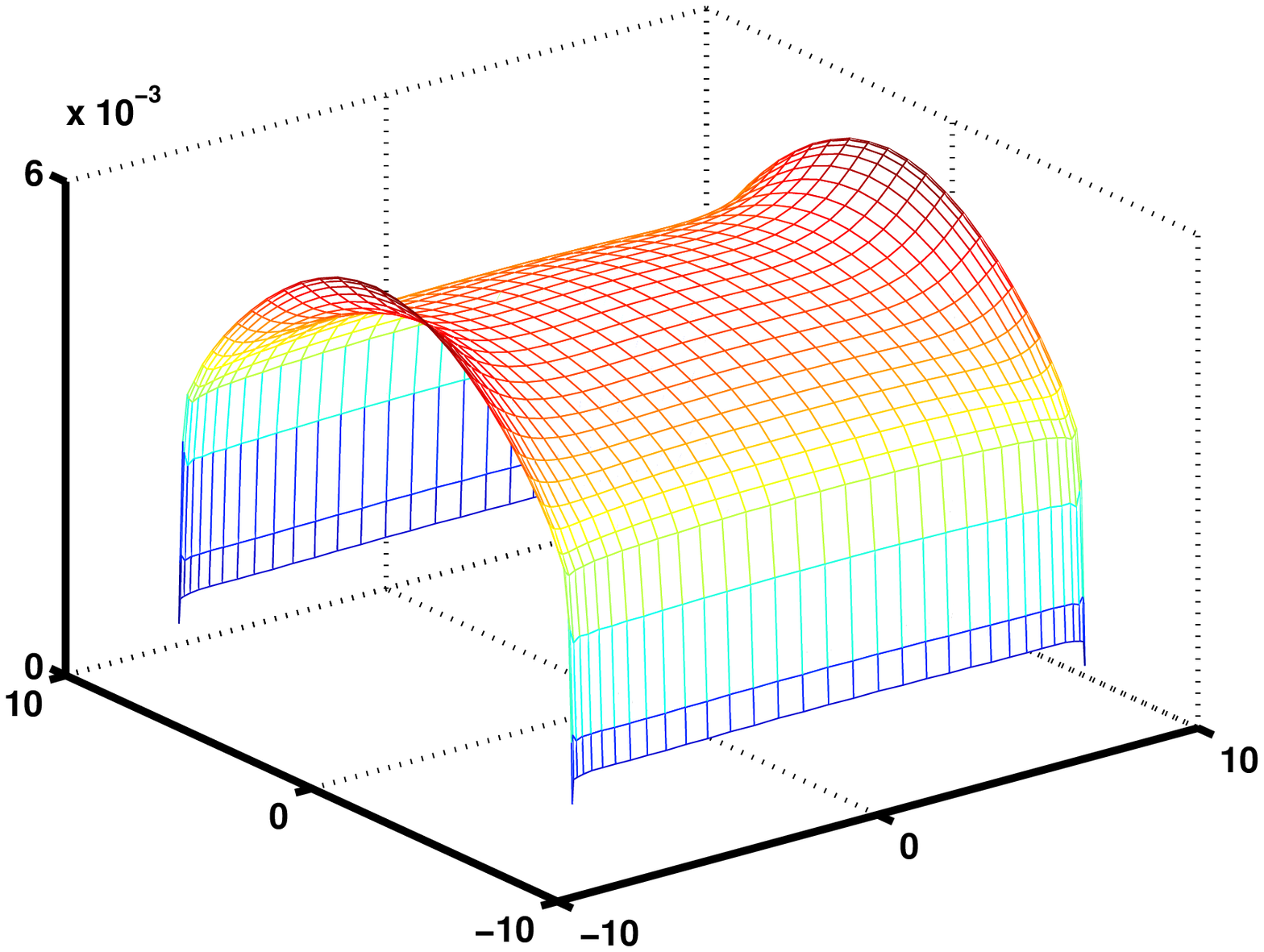}
\includegraphics[width=6.0cm,height=6.0cm]{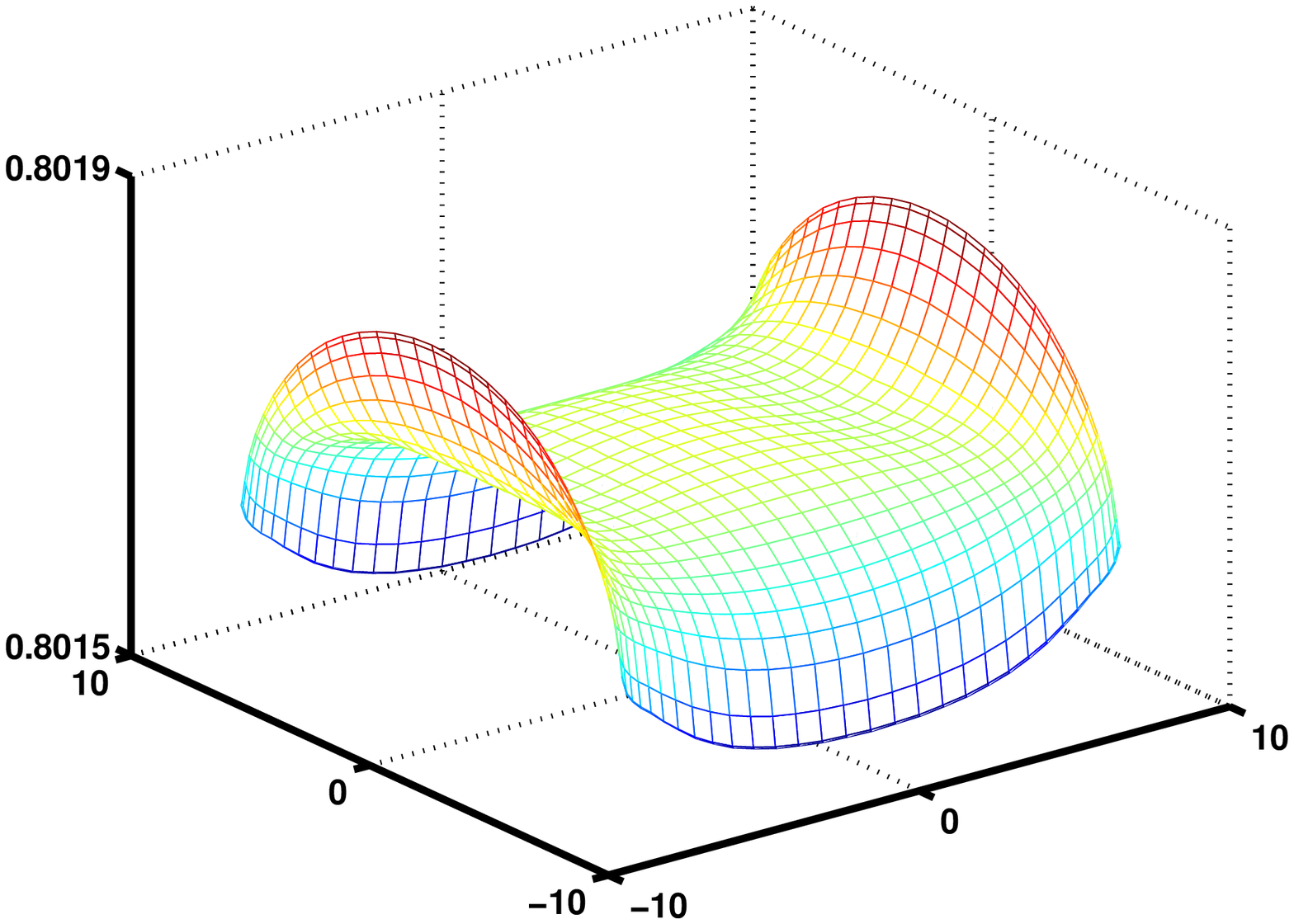}\\[0pt]($\mathbf{A}$)
\hspace{50mm} ($\mathbf{B}$)\\[0pt]%
\includegraphics[width=6.0cm,height=6.0cm]{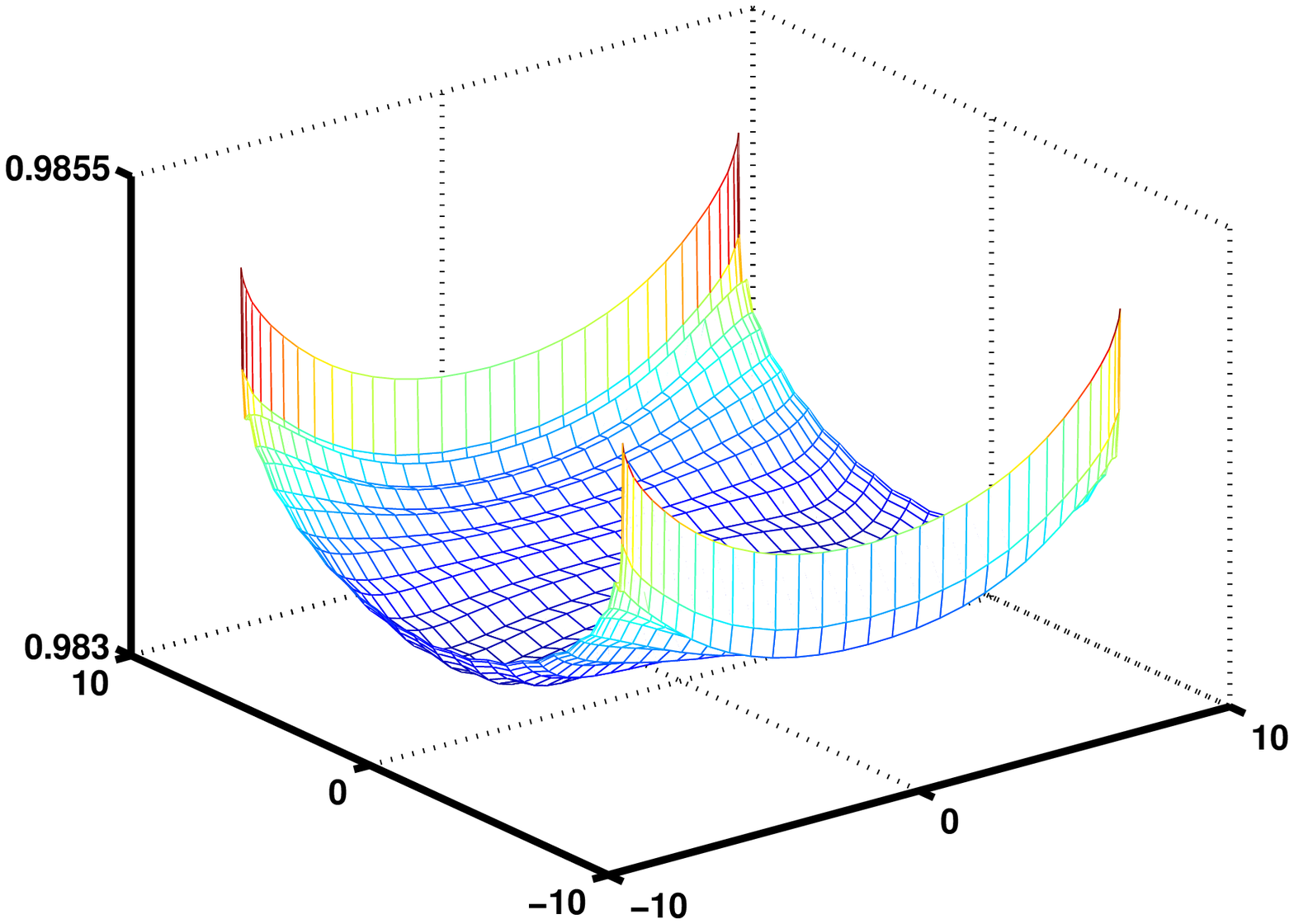}\\
$\mathbf{C}$
\end{center}
\caption{{\protect\small The order parameter (S) distribution for the top ($\mathbf{A}$), middle ($\mathbf{B}$)
and bottom ($\mathbf{C}$) horizontal slices of the LCEs sample at the equilibrium state. The order
parameter is close to zero on the top slice while it is close to one on the bottom
slice. It is the nonhomogeneous distribution of the order parameter that leads to
internal stress, and thus results in the shape changes of the LCEs sample. Moreover,
the order parameter slightly varies on each of these slices, suggesting the elastic
effect on the order parameter. }}%
\label{Saddle-S}%
\end{figure}

The simulation shows that as the dynamics progresses, the order
parameter in each horizontal layer converges to nearly constant values
essentially determined by the temperature assigned to that layer (see
Eq.~(\ref{PDE-S})), though somewhat affected also by elastic effects
induced by coupling to $\mathbf{n}$.  This is illustrated in Figure
\ref{Saddle-S}, which shows that the order parameter generally assume
smaller values at the top and larger values on the bottom, but also
varies (slightly) within each layer.

\begin{figure}[h]
\begin{center}
\includegraphics[width=6.0cm,height=6.0cm]{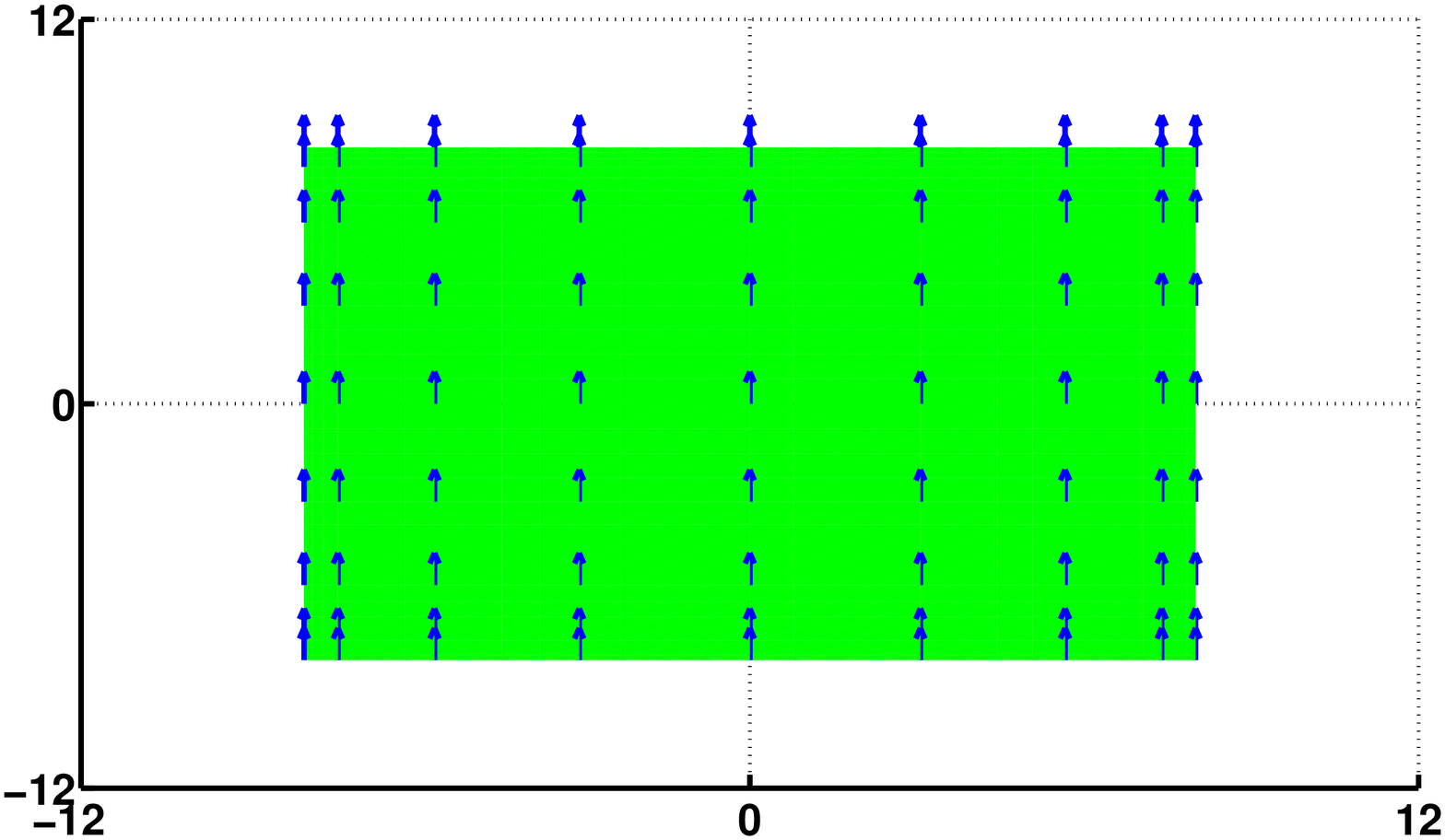}\\
($\mathbf{A}$)\\
\includegraphics[width=6.0cm,height=6.0cm]{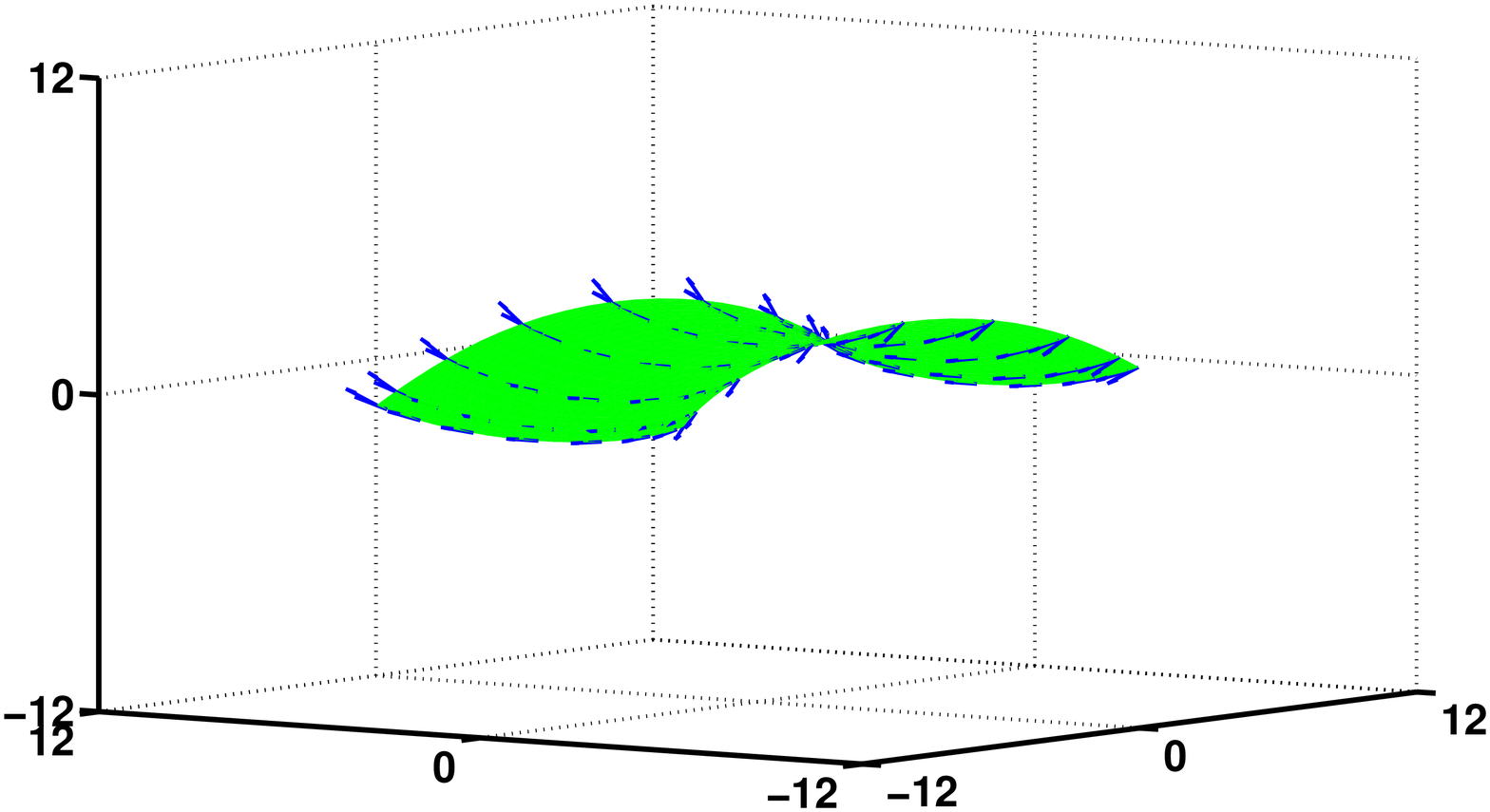}
\includegraphics[width=6.0cm,height=6.0cm]{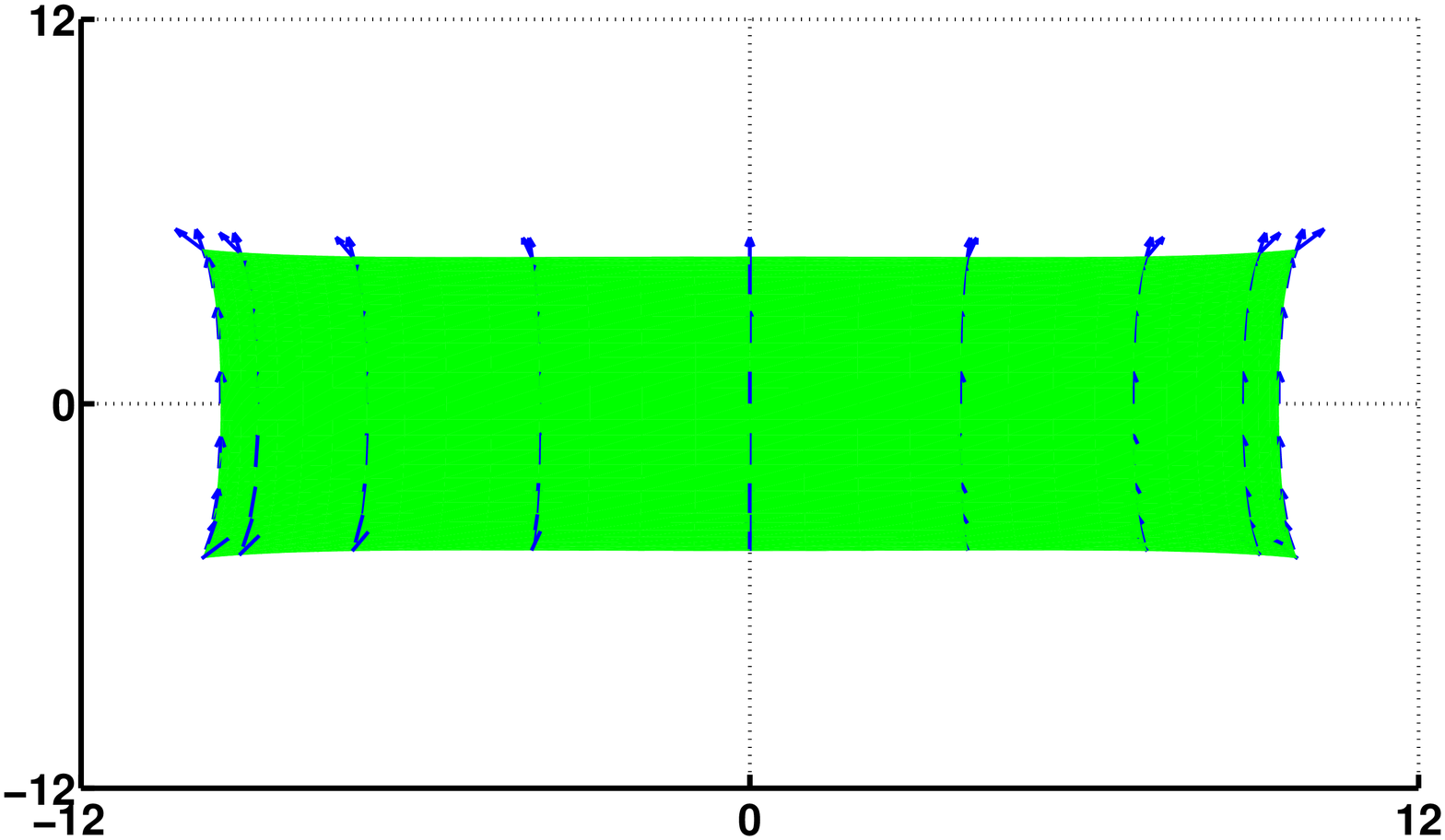}\\
($\mathbf{B}$) \hspace{50mm} ($\mathbf{C}$)\\
\includegraphics[width=6.0cm,height=6.0cm]{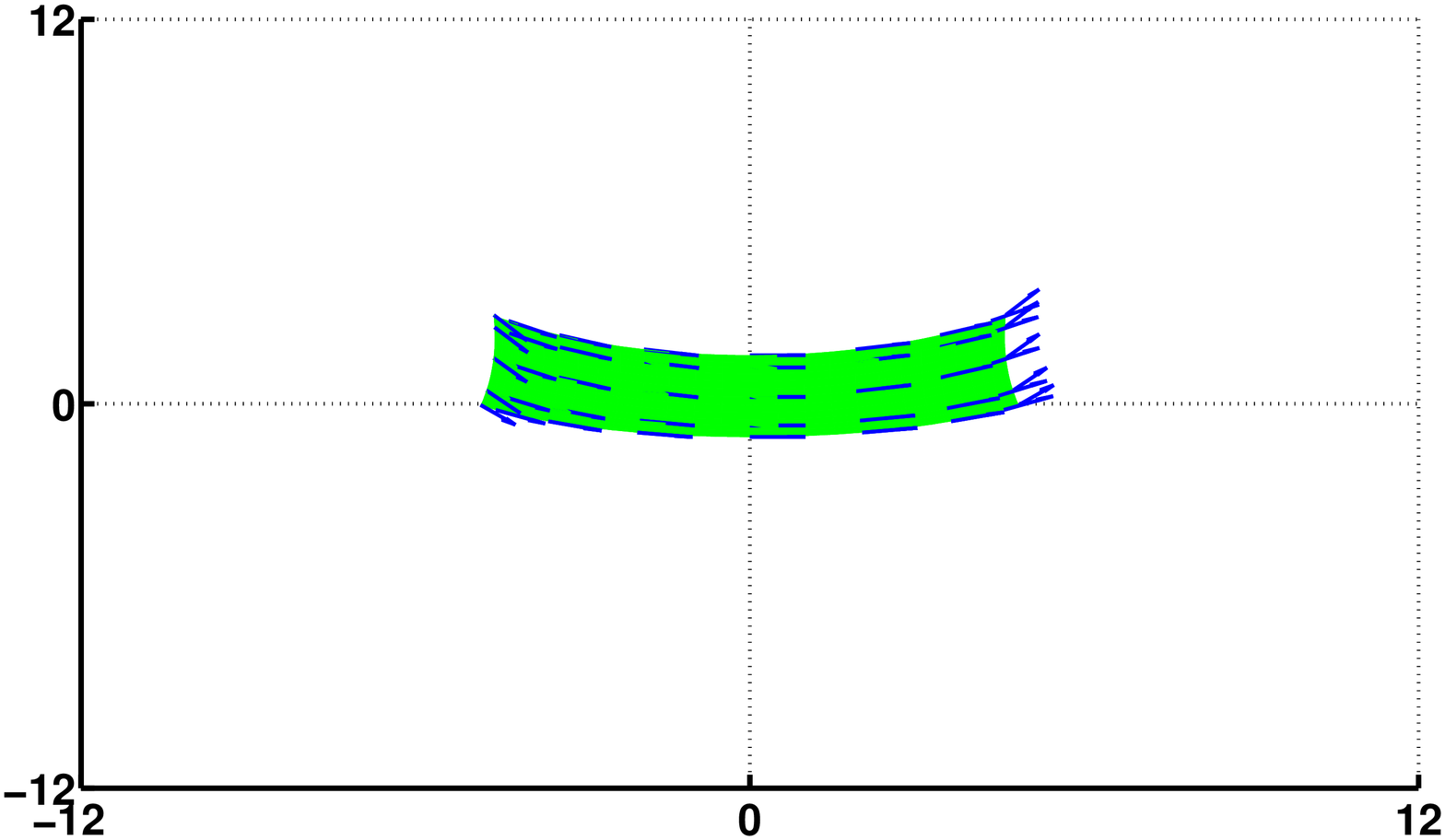}\\
($\mathbf{D}$)
\end{center}
\caption{{\protect\small The nematic direction ($\mathbf{n}$) distribution for the top slice of the LCEs
sample at the initial and the equilibrium state. The plot ($\mathbf{A}$) represents the nematic direction
on the top slice at the initial state. The plots ($\mathbf{B}$), ($\mathbf{C}$) and ($\mathbf{D}$)
illustrate the nematic direction on the top slice for different perspectives at the equilibrium state. }}%
\label{Saddle-N}%
\end{figure}

We also study the dynamics of nematic director $\mathbf{n}$. In Figure
\ref{Saddle-N}, the nematic director on the top surface of the sample is
compared at the initial and equilibrium states.  For the equilibrium
state, three perspectives are given from which one can easily discern
the evolution of the nematic director. Similar director distributions
on the other layers of the sample can be observed.

The dynamics of the second simulation can be explored similarly. In
this simulation, all initial conditions and spatial temperature
distributions are as in the first example, except that one lateral
surface of the sample is fixed, and the dimensions of the LCE sample
are now $4\times8\times1$, which is narrower in the $x-$direction.

\begin{figure}[h]
\begin{center}
\includegraphics[width=5.5cm,height=5.5cm]{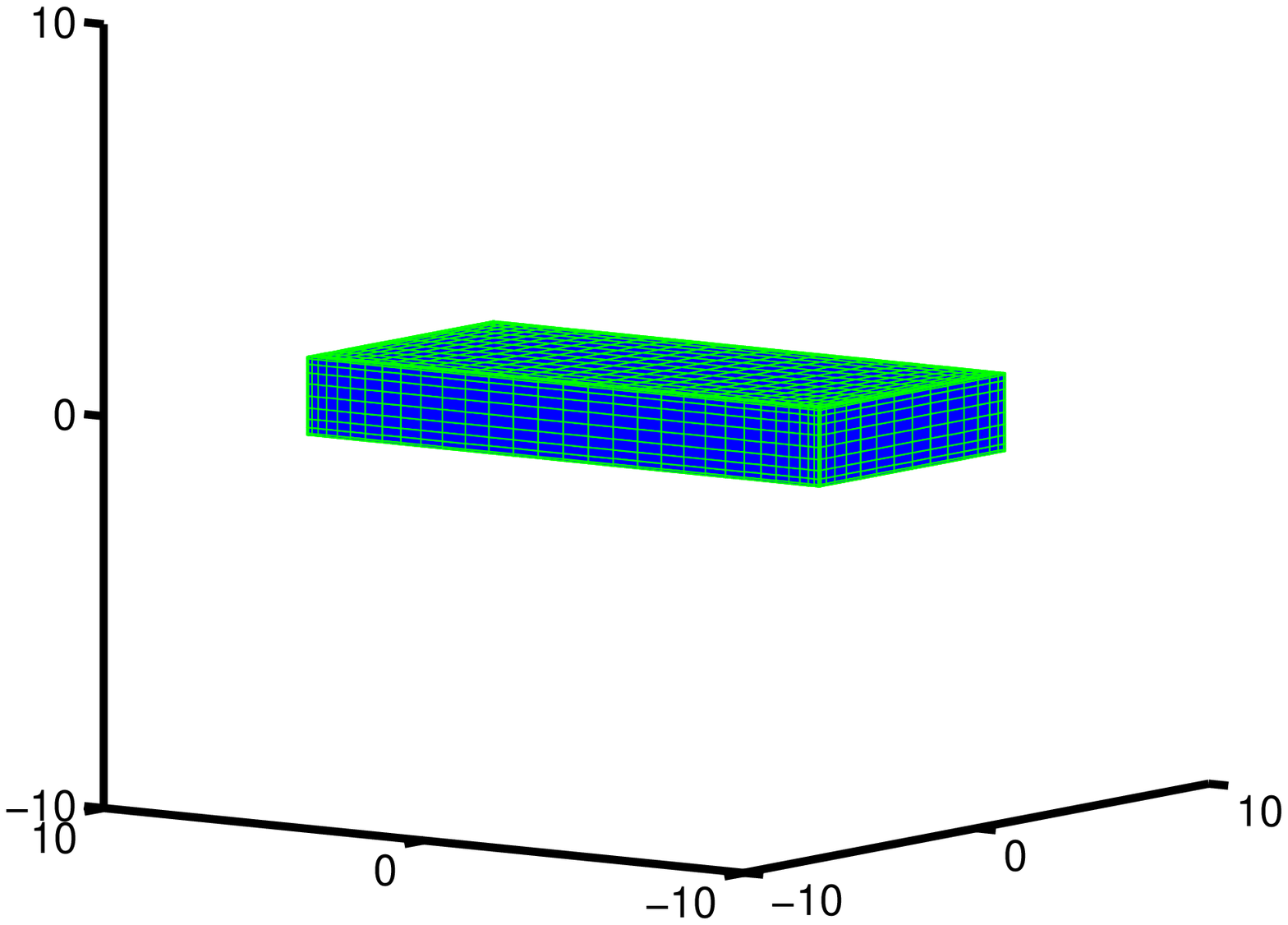}
\includegraphics[width=5.5cm,height=5.5cm]{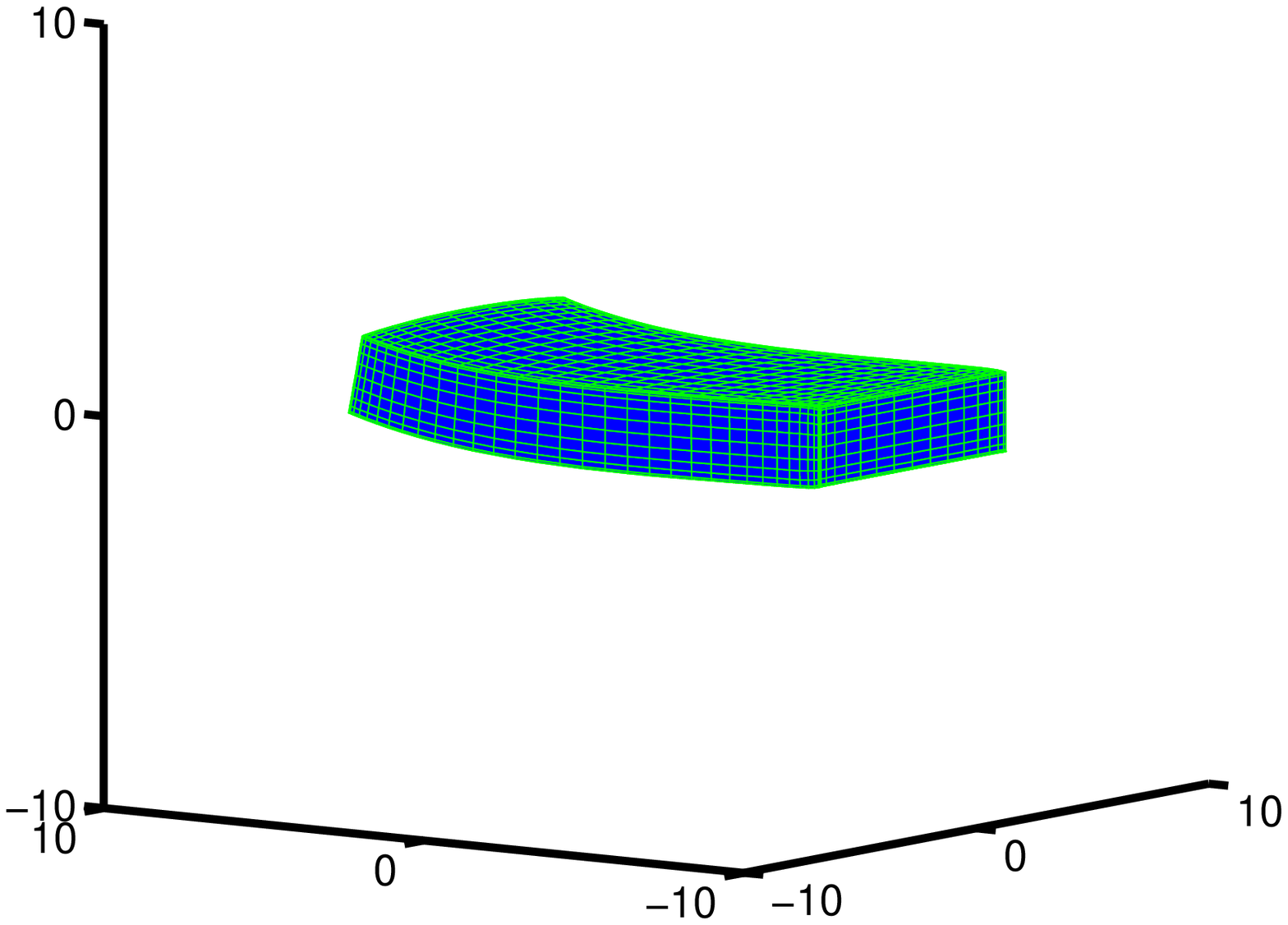}\\[0pt]($\mathbf{A}$)
\hspace{50mm} ($\mathbf{B}$)\\[0pt]%
\includegraphics[width=5.5cm,height=5.5cm]{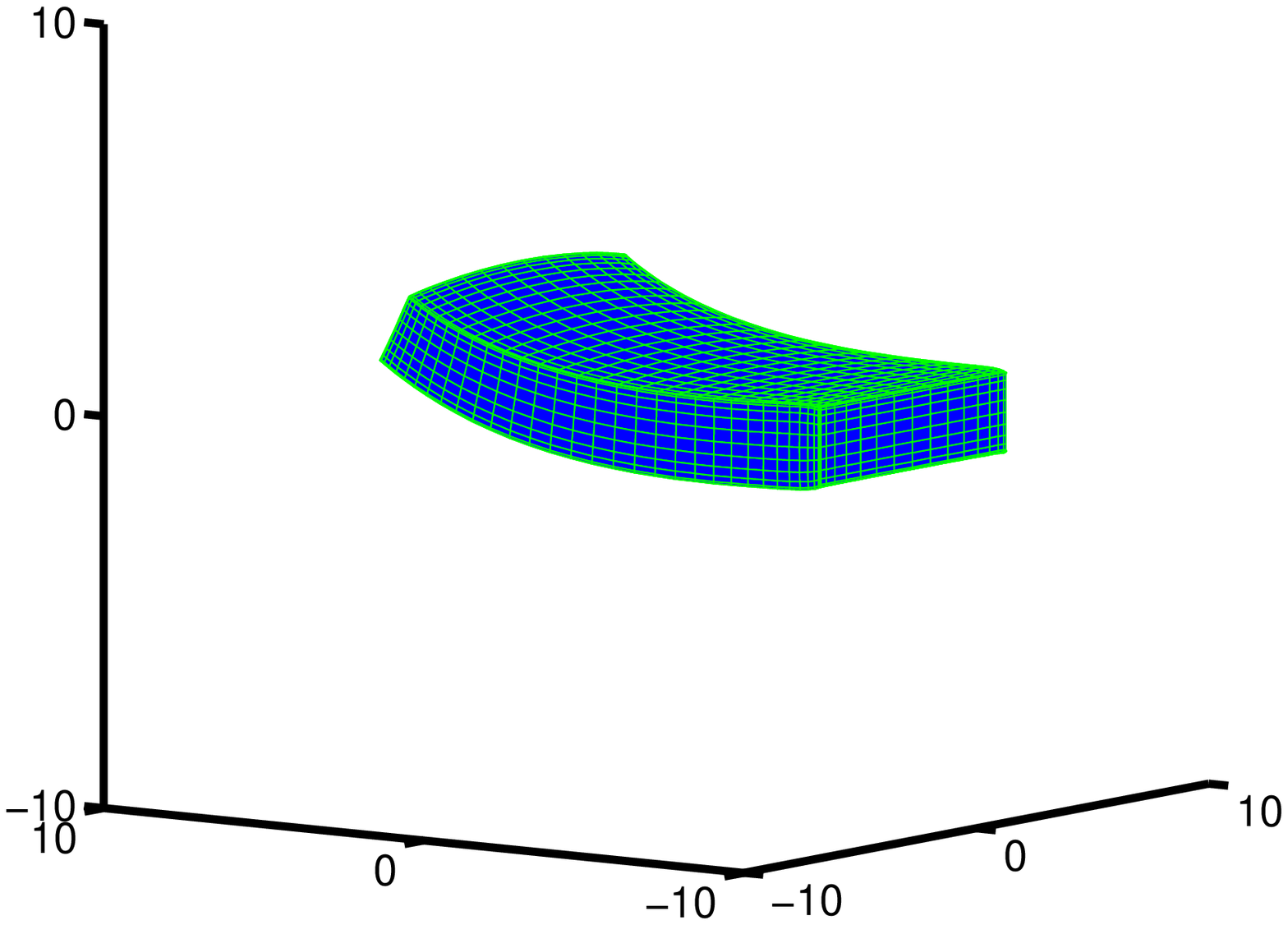}
\includegraphics[width=5.5cm,height=5.5cm]{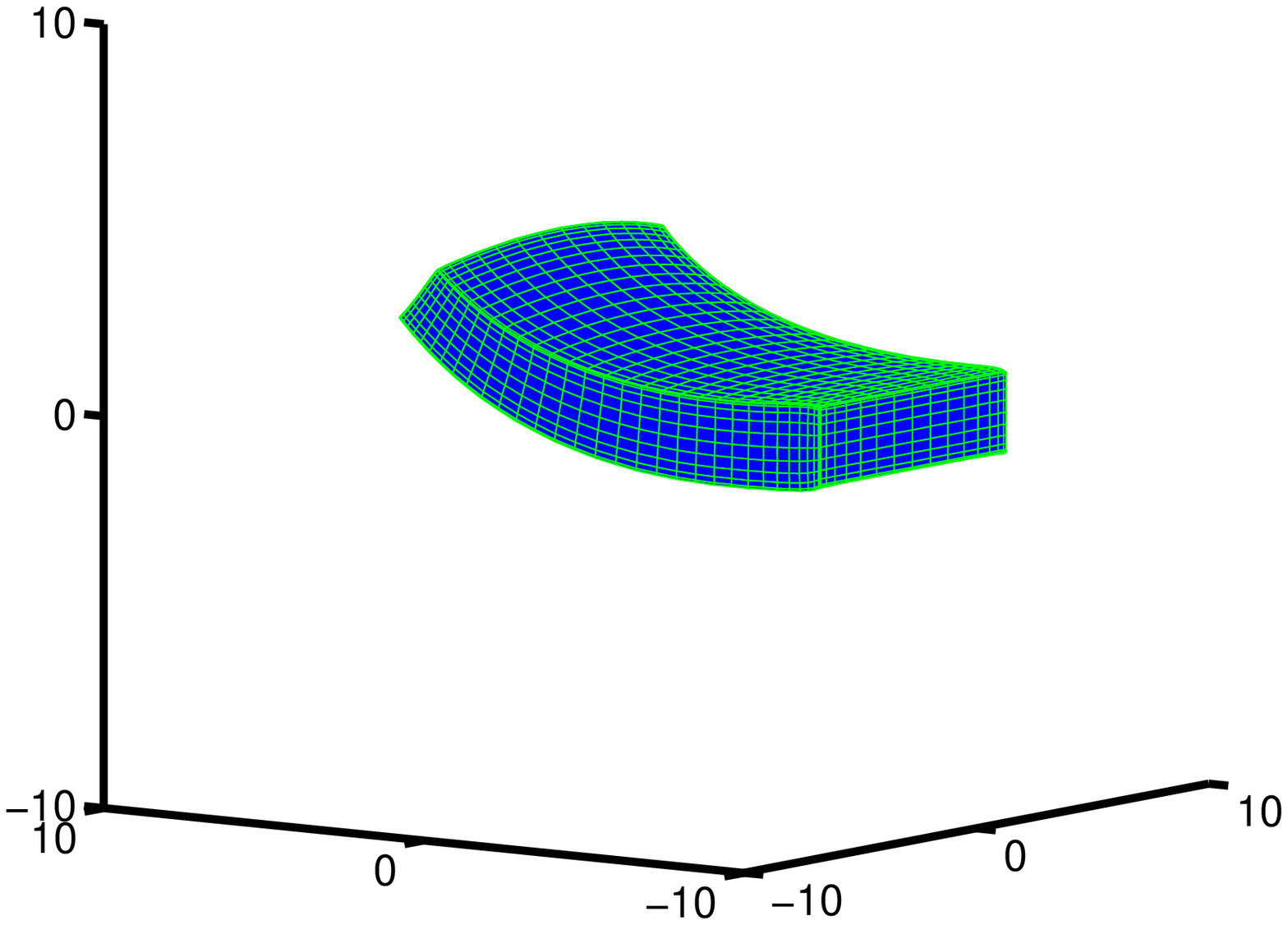}\\[0pt]($\mathbf{C}$)
\hspace{50mm} ($\mathbf{D}$)\\[0pt]%
\includegraphics[width=5.5cm,height=5.5cm]{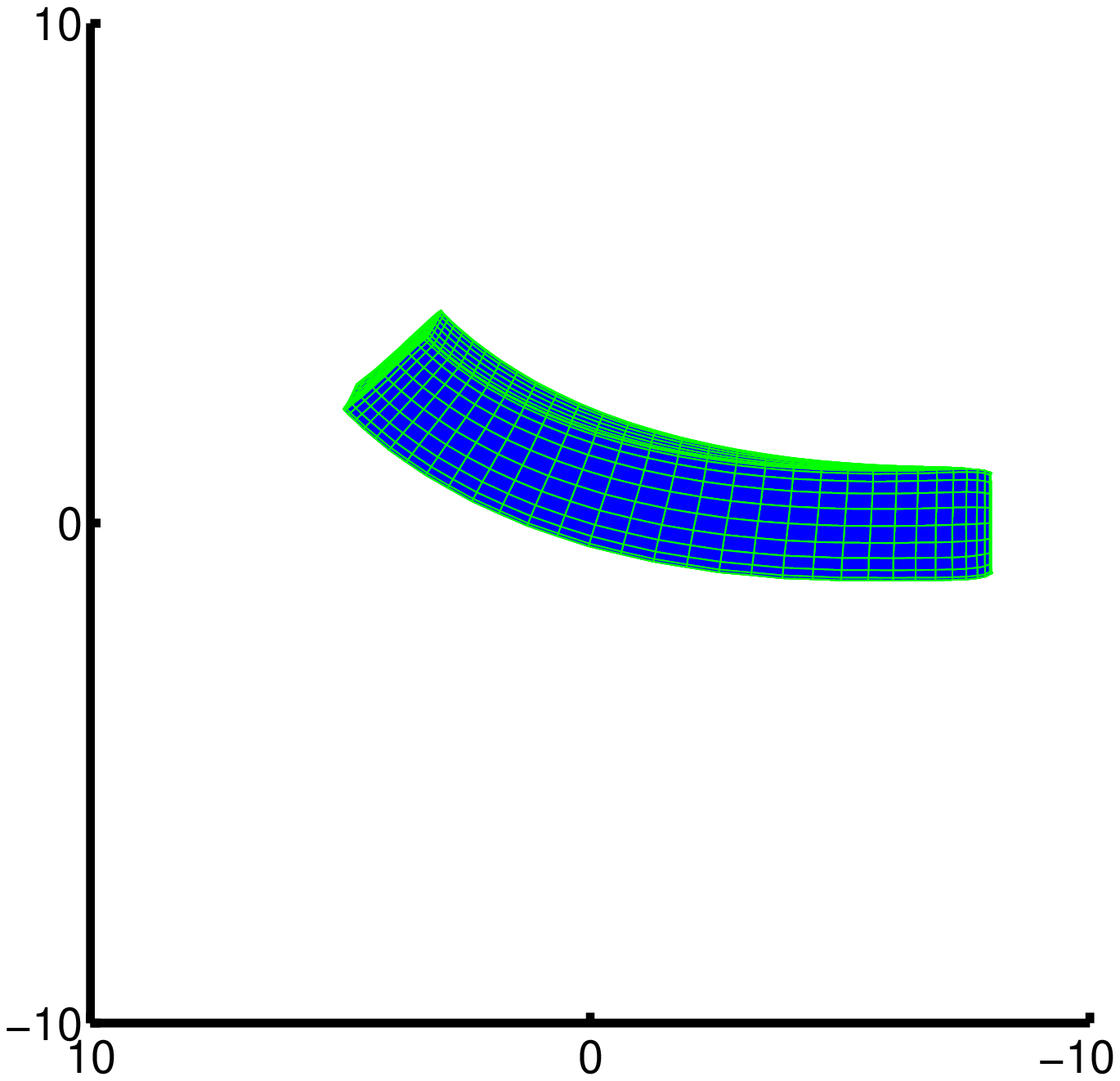}
\includegraphics[width=5.5cm,height=5.5cm]{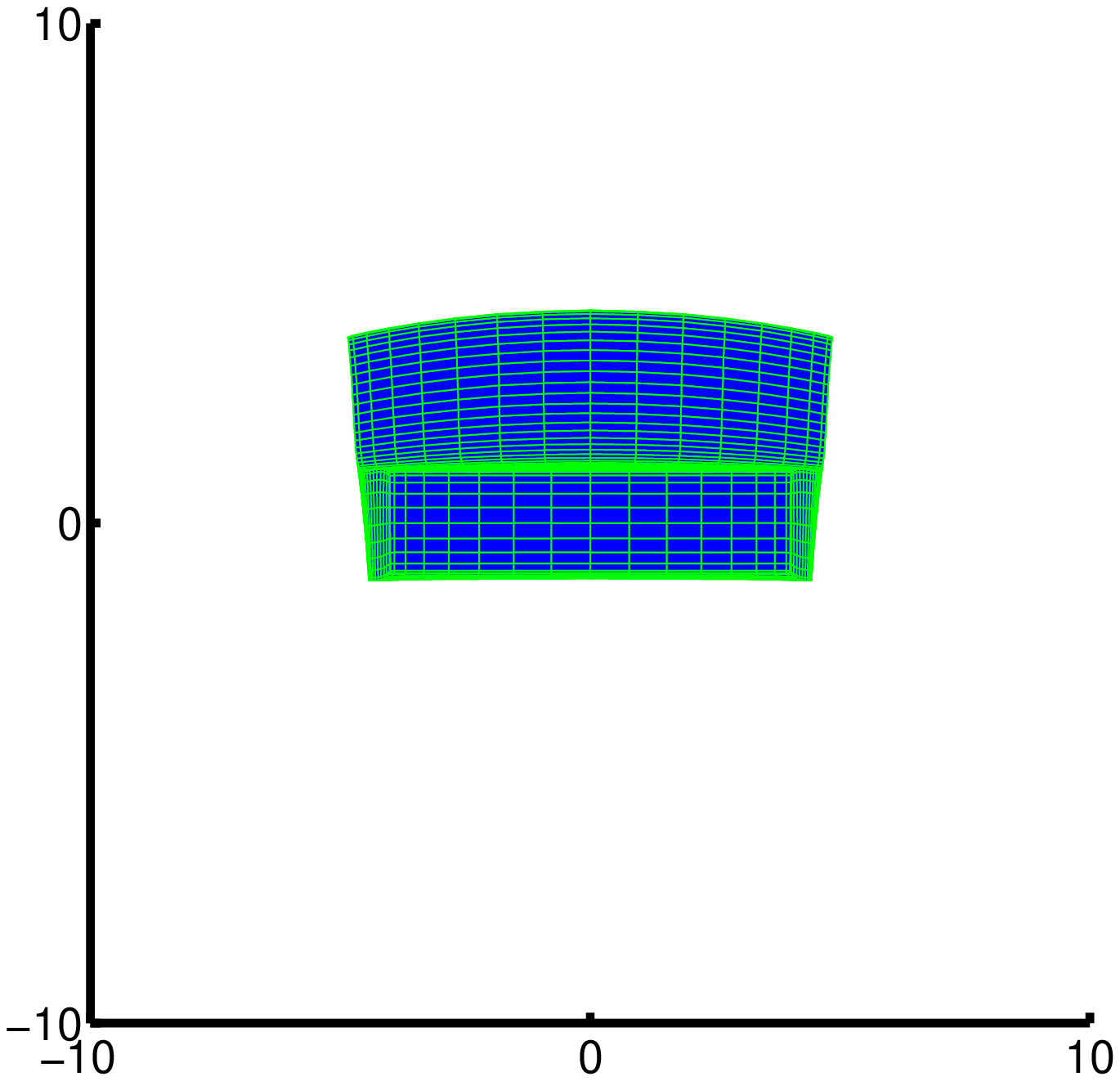}\\[0pt]($\mathbf{E}$)
\hspace{50mm} ($\mathbf{F}$)\\[0pt]
\end{center}
\caption{{\protect\small The shape evolution of the LCEs sample due to nonhomogeneous changes in temperature.
Figure $\mathbf{A}$ shows the initial state of the LCEs sample, while Figures $\mathbf{B}$, $\mathbf{C}$ are two intermediate states and
Figure $\mathbf{D}$ represents the equilibrium state. Figures $\mathbf{E}$ and $\mathbf{F}$ present the shape of the LCEs sample at the
equilibrium state (Figure $\mathbf{D}$) from two different perspectives. This experiment shares the same condition
as the previous simulation in Figure \ref{Saddle-shape} except that a lateral surface of the LCEs sample
is fixed. This numerical experiment simulates the real one shown in Figure 2 in \cite{PCFS04}.}}%
\label{Bending-shape}%
\end{figure}

Figure \ref{Bending-shape} shows the deformation process. Again, the
result is very similar to that observed in actual experiment (see
Fig.~2 of \cite{PCFS04}) with the sample bending upwards at its free
end.  As the in first simulation, the deformation proceeds through
three stages, with the sample also contracting along the initial
nematic direction and extending in the other two orthogonal
directions. This is illustrated the Figs.~\ref{Bending-shape}E and F.
All these phenomena share the underlying physics as in the first
simulation.

\begin{figure}[h]
\begin{center}
\includegraphics[width=6.0cm,height=6.0cm]{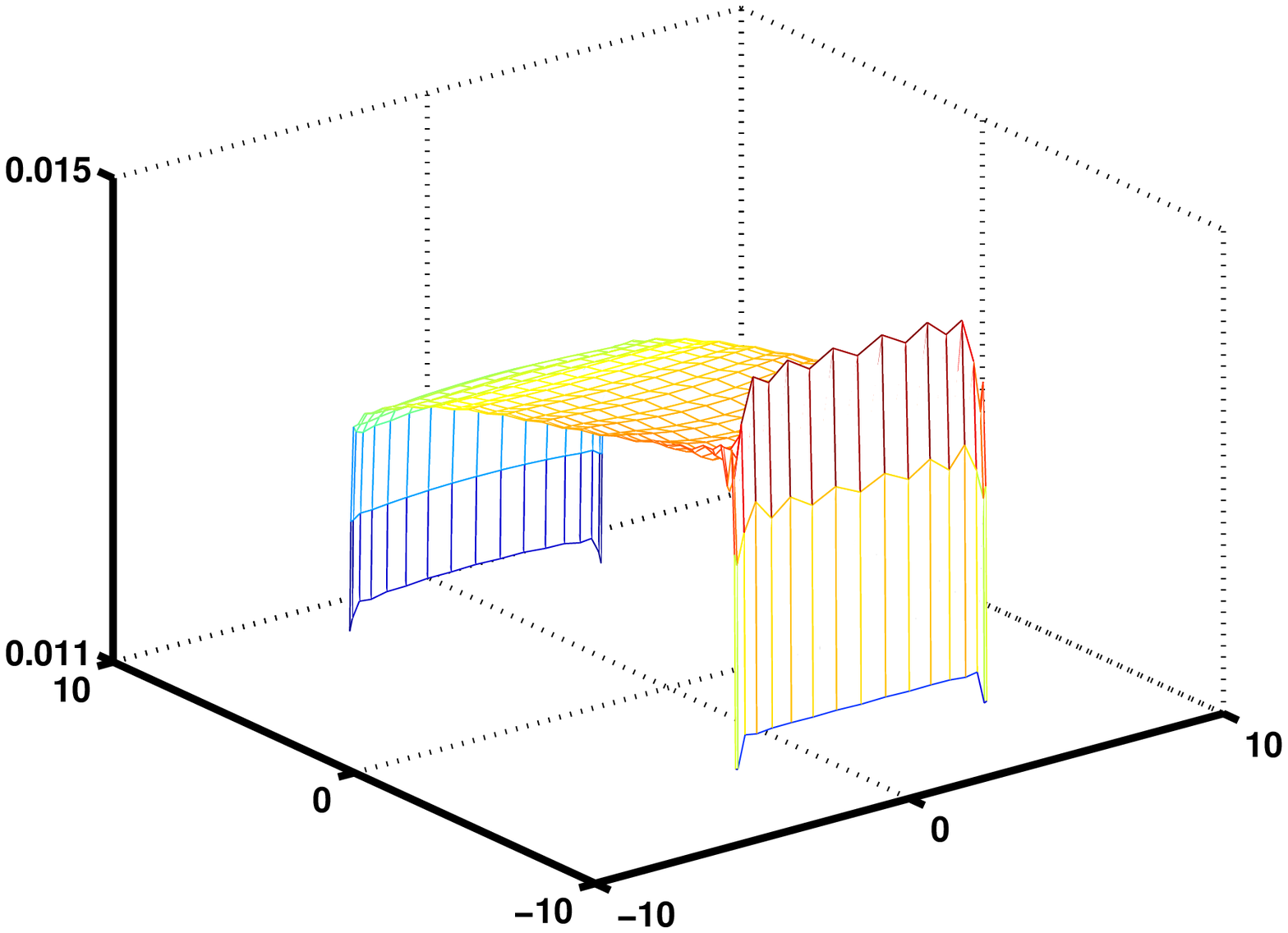}
\includegraphics[width=6.0cm,height=6.0cm]{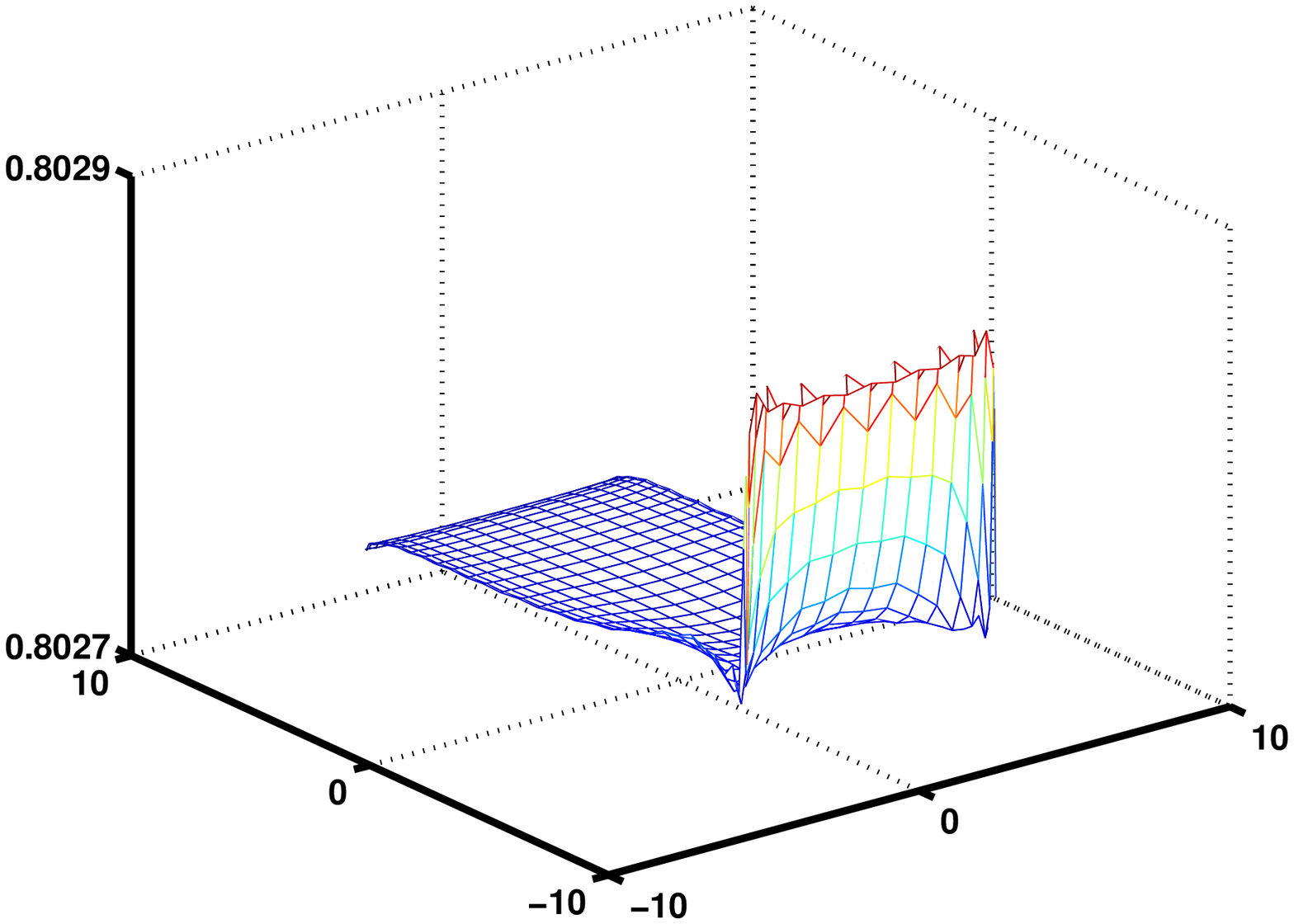}\\[0pt]($\mathbf{A}$)
\hspace{50mm} ($\mathbf{B}$)\\[0pt]%
\includegraphics[width=6.0cm,height=6.0cm]{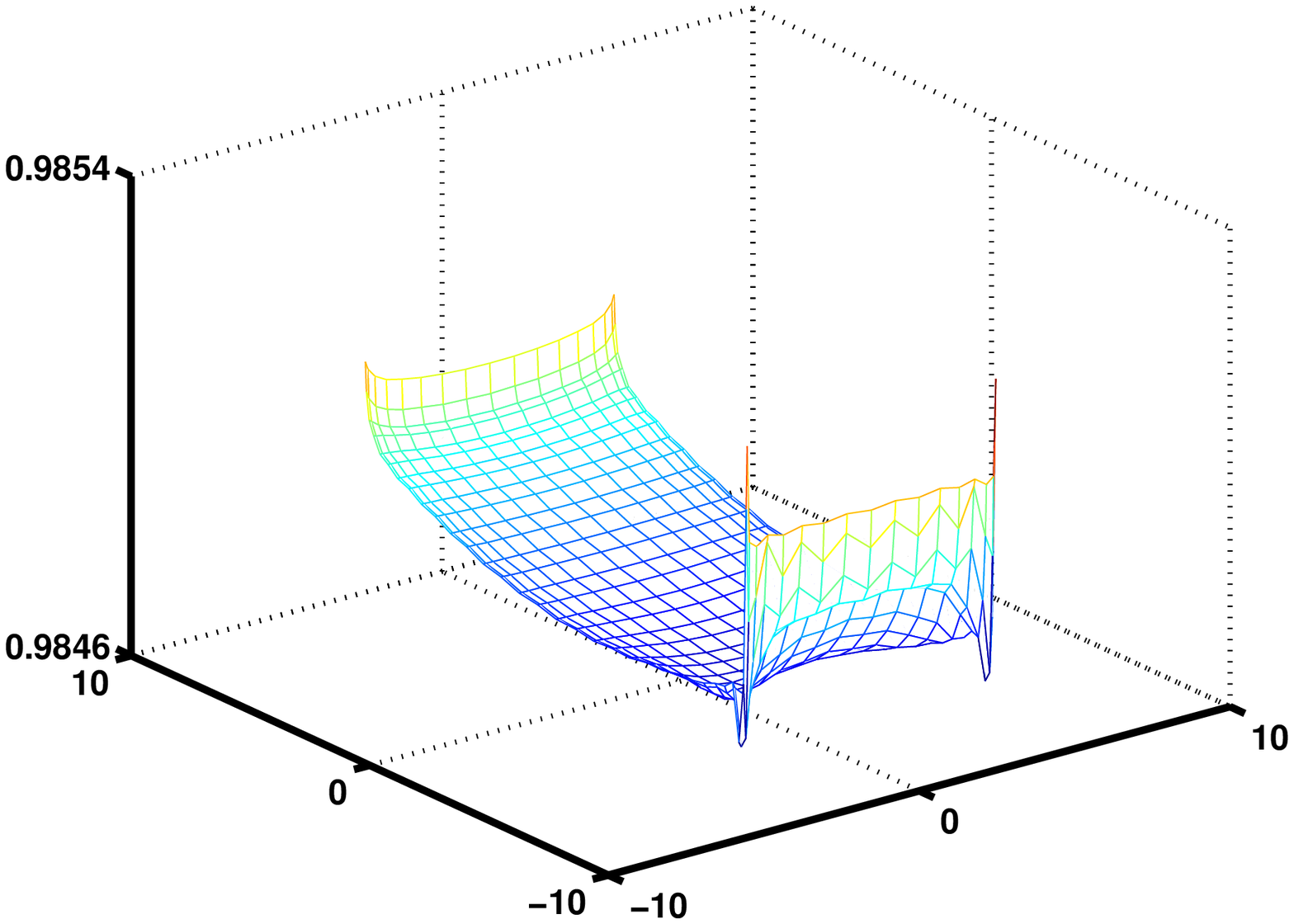}\\
$\mathbf{C}$
\end{center}
\caption{{\protect\small The order parameter (S) distribution for the top ($\mathbf{A}$), middle ($\mathbf{B}$)
and bottom ($\mathbf{C}$) horizontal slices of the LCEs sample at the equilibrium state. The order
parameter is close to zero on the top slice while it is close to one on the bottom
slice. It is the nonhomogeneous distribution of the order parameter that leads to
internal stress, and thus results in the shape changes of the LCEs sample. Moreover,
the order parameter slightly varies on each of these slices and oscillates near the fixed lateral surface
of the LCEs sample, suggesting both the elastic effect and the effect due to the anchored surface
on the order parameter. }}%
\label{Bending-S}%
\end{figure}

In Figure \ref{Bending-S}, the order parameter distributions within
the top, middle, and bottom layers of the sample in the equilibrium
state are shown.  The order parameter within each layer is nearly
constant within each layer, though with small oscillations caused by
boundary effects. However, when compared with Fig.~\ref{Saddle-S} for
the first simulation, one finds that the basic deformation pattern is
different and asymmetric due to the anchoring boundary condition used
in this experiment.

\begin{figure}[h]
\begin{center}
\includegraphics[width=6.0cm,height=6.0cm]{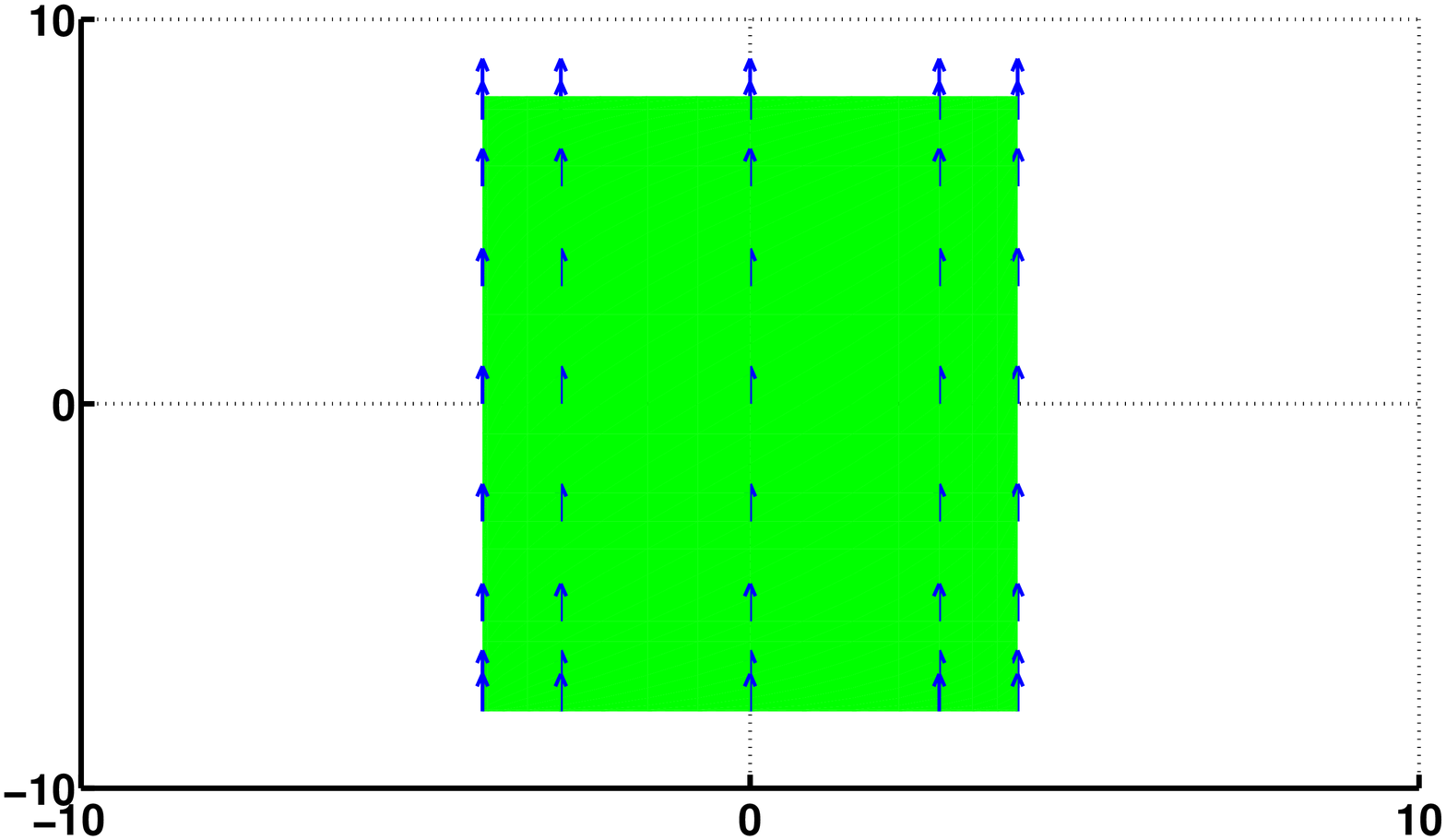}\\
($\mathbf{A}$)\\
\includegraphics[width=6.0cm,height=6.0cm]{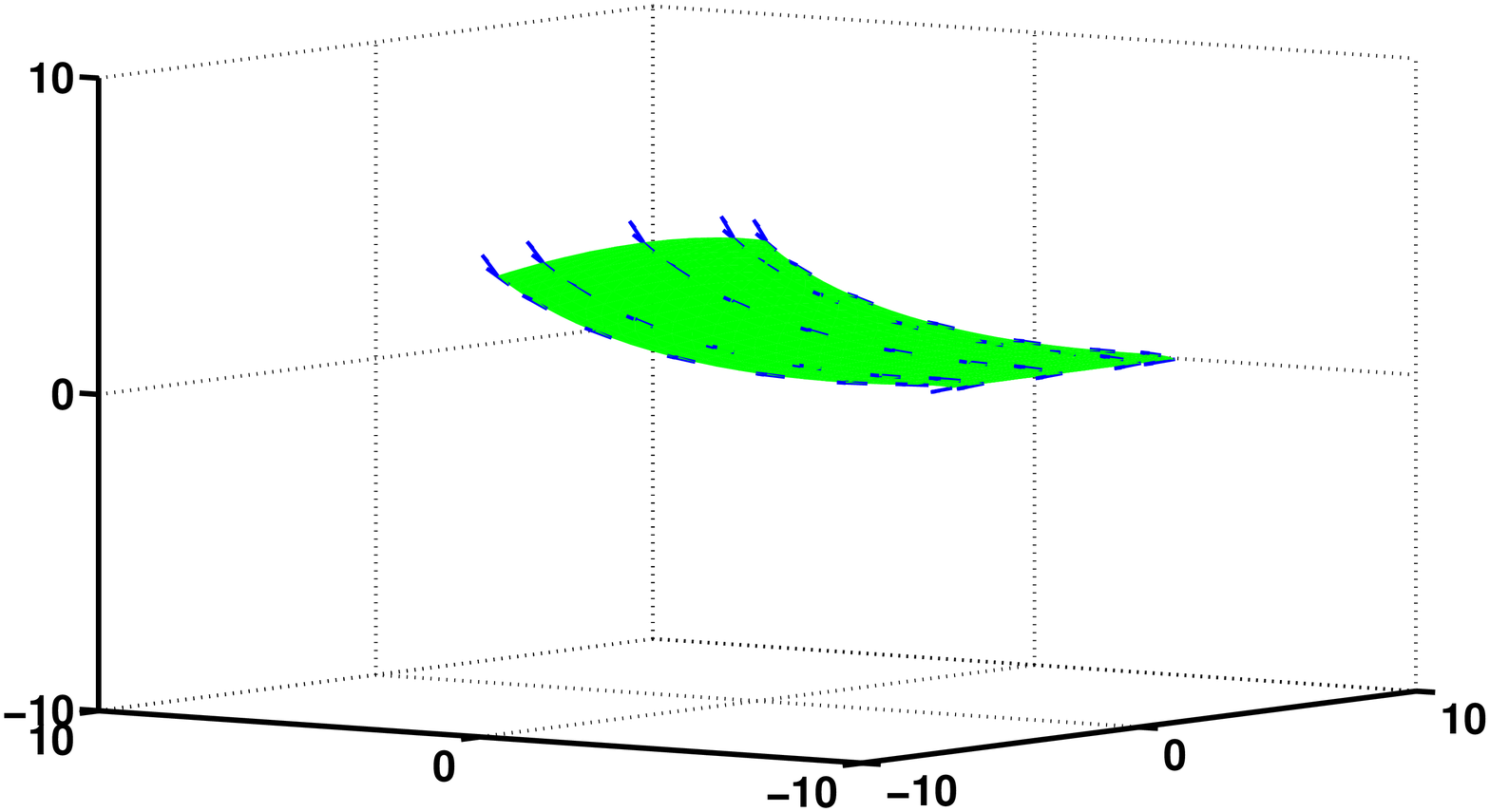}
\includegraphics[width=6.0cm,height=6.0cm]{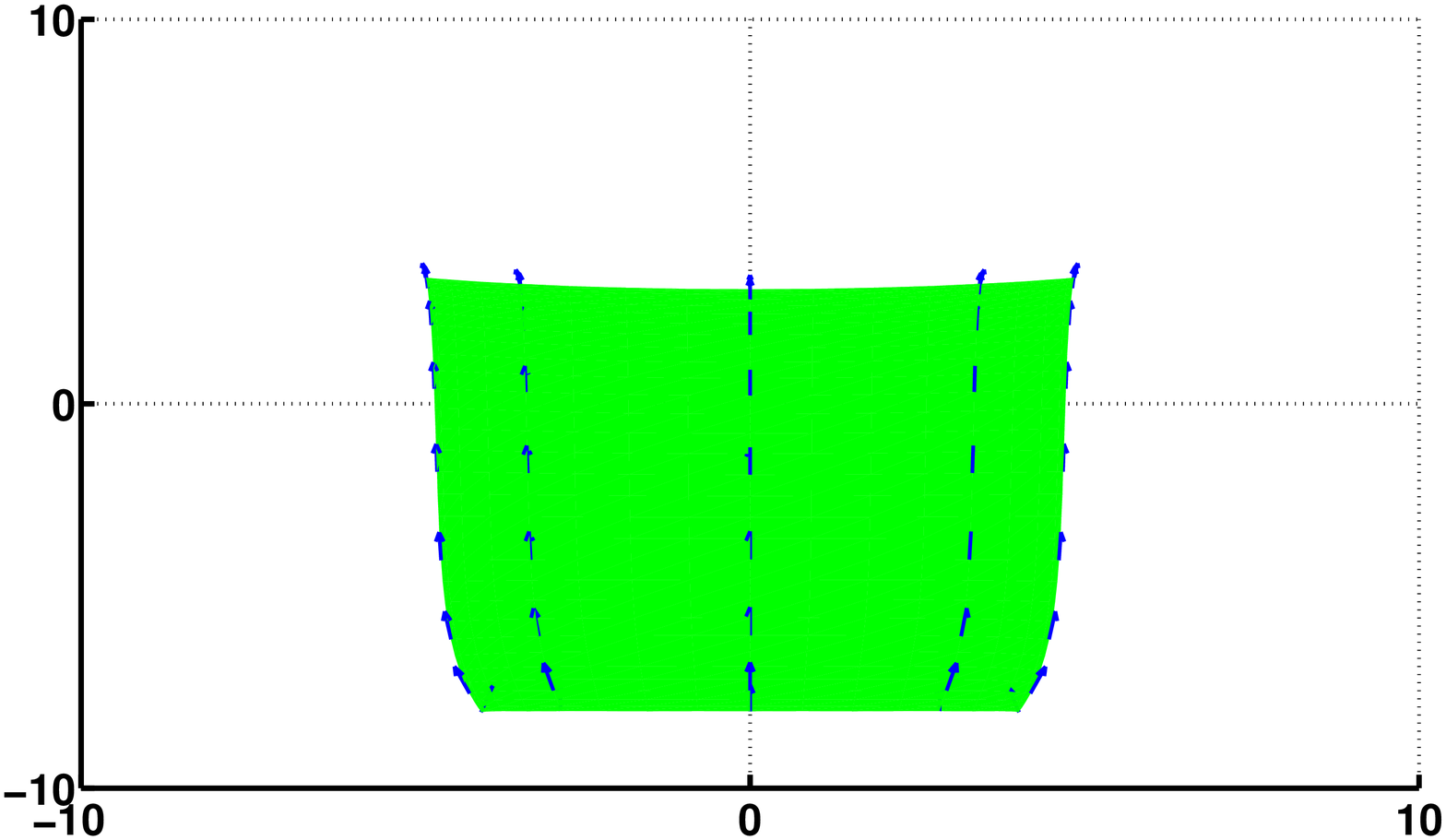}\\
($\mathbf{B}$) \hspace{50mm} ($\mathbf{C}$)\\
\includegraphics[width=6.0cm,height=6.0cm]{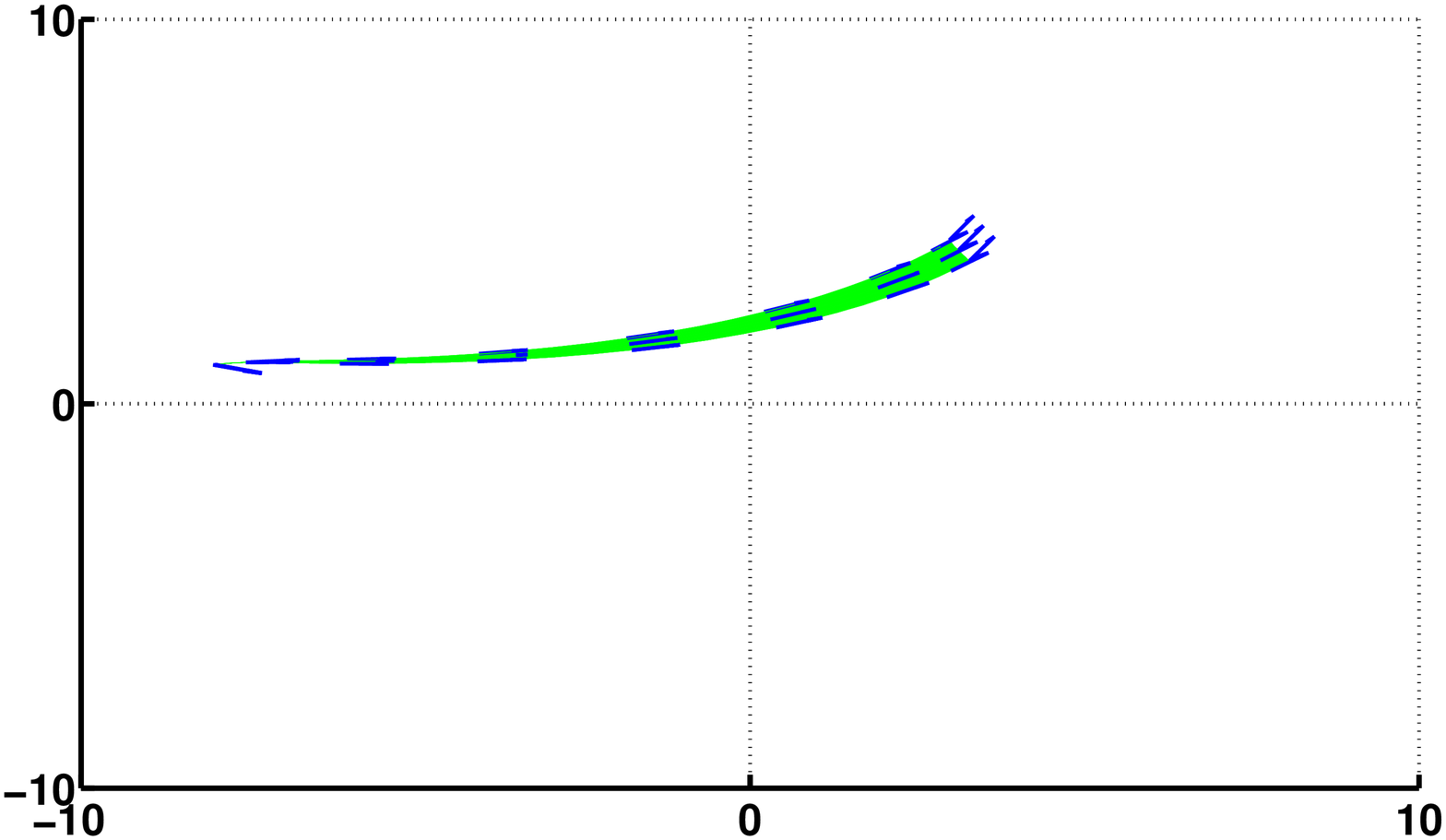}\\
($\mathbf{D}$)
\end{center}
\caption{{\protect\small The nematic direction ($\mathbf{n}$) distribution for the top slice of the LCEs
sample at the initial and the equilibrium state. The plot ($\mathbf{A}$) represents the nematic direction
on the top slice at the initial state. The plots ($\mathbf{B}$), ($\mathbf{C}$) and ($\mathbf{D}$)
illustrate the nematic direction on the top slice for different perspectives at the equilibrium state. }}%
\label{Bending-N}%
\end{figure}

Figure \ref{Bending-N}, shows the disposition of the nematic director on the top
surface of the LCEs sample, again comparing the initial and the equilibrium
states. For the equilibrium state, three perspectives are shown. We see that
the nematic director bends upwards, and the bending increases with distance
from the fixed side of the sample. Similar director dynamics are observed
in the other layers.

\section{Conclusions}

In this paper, we derived the equations of motion for an LCE sample
in the long-wave limit, and implemented their numerical solution. Our numerical
experiments demonstrate that the model is capable of describing the dynamics
of nematic LCEs when exposed to external stimuli such as illumination.

\section{Acknowledgements}

The authors acknowledge the support of the NSF and DOE.

\end{document}